\PassOptionsToPackage{pdfpagelabels=false}{hyperref}
\documentclass[a4paper,fleqn,usenatbib]{mnras}

\usepackage[T1]{fontenc}
\usepackage{ae,aecompl}
\usepackage{adjustbox}
\usepackage{multicol}        % Multi-column entries in tables
\usepackage{multirow}        % Multi-row entries in tables

\usepackage{epsfig,amstext,alltt,fancyhdr,setspace}
\usepackage{graphicx}
\usepackage{lmodern} %correct
\usepackage{color}
\usepackage{ucs}
\usepackage[utf8x]{inputenc}
\usepackage{url}
\usepackage{mathrsfs}
\usepackage{amsmath} %correct
\usepackage{amssymb}
\usepackage{appendix}
\usepackage{rjwmath}
\usepackage{varioref}
\usepackage{times}
\usepackage{enumerate}
\usepackage{magazin_abbrev}
\def \Eqt{Eq.\thinspace}
\def \sect{Sect.\thinspace}
\def \fig{Fig.\thinspace}
\def \tab{Table\thinspace}
\def \App{Appendix.\thinspace}

\def\P{{\cal P}}

\def\N{{\cal N}}

\def\d{{\rm d}}

\def\CMb{\boldsymbol{\tens{C}}}
\def\HMb{\boldsymbol{\tens{H}}}

\title[Revisiting CFHTLenS with COSEBIs and CCOSEBIs]
{Revisiting CFHTLenS cosmic shear:  Optimal E/B mode decomposition using COSEBIs and compressed COSEBIs}
\author[M. Asgari et al.]{
Marika Asgari,$^{1}$\thanks{E-mail: ma@roe.ac.uk}
Catherine Heymans$^{1}$,
Chris Blake$^{2}$,
Joachim Harnois-Deraps$^{3}$, \newauthor \;Peter Schneider$^{4}$,
Ludovic Van Waerbeke$^{3}$
\\
$^{1}$Scottish Universities Physics Alliance, Institute for Astronomy, University of Edinburgh, Royal Observatory,
Blackford Hill, Edinburgh,\\ \; EH9 3HJ, U.K.\\
$^{2}$Centre for Astrophysics \& Supercomputing, Swinburne University of Technology, P.O Box 218, Hawthorn, VIC 3122, Australia.\\
$^{3}$Department of Physics and Astronomy, University of British Columbia, 6224 Agricultural Road, Vancouver, V7T 1Z1, B.C, Canada.\\
$^{4}$Argelander-Institut f\"ur Astronomie, Universit\"at Bonn, Auf dem H\"ugel 71, D-53121 Bonn, Germany             
}

\date{Accepted 2016 October 4. Received 2016 August 30; in original form 2015 December 21}

\pubyear{2017}

\begin{document}

\label{firstpage}
\pagerange{\pageref{firstpage}--\pageref{lastpage}}
\maketitle

% Abstract of the paper
\begin{abstract}
% TBD - will circulate via e-mail asap
We present a re-analysis of the CFHTLenS weak gravitational lensing survey 
using Complete Orthogonal Sets of E/B-mode Integrals, known as COSEBIs.
COSEBIs provide a complete set of functions to efficiently separate 
E-modes from B-modes and hence allow for robust and stringent tests for 
systematic errors in the data.  This analysis reveals significant 
B-modes on large angular scales that were not previously seen using the standard
E/B decomposition analyses.  We find that the significance of the 
B-modes is enhanced when the data is split by galaxy type and 
analysed in tomographic redshift bins.   
Adding tomographic bins to the analysis increases the number of COSEBIs modes, 
which results in a less accurate estimation of the covariance matrix 
from a set of simulations.   We therefore also present the first compressed COSEBIs 
analysis of survey data, where the COSEBIs modes are optimally
combined based on their sensitivity to cosmological parameters.  
In this tomographic CCOSEBIs analysis we find the B-modes to be consistent with zero when
the full range of angular scales are considered.   

%Cosmic shear analysis is a powerful probe of the geometry of the Universe as well as structure evolution. 
%Accurate cosmic shear analysis depends on many factors from data reduction to statistical methods used. 
%COSEBIs is a method that fully separates E-modes from B-modes for a finite angular range, hence it is a great test of systematics. 
%We use COSEBIs to measure the cosmic shear signal in CFHTLenS data and show the first tomographic COSEBIs measurements.
%The tension between the CFHTLenS and Planck cosmological results have been widely reported. 
%We use COSEBIs to further investigate the significance of systematic errors in CFHTLenS which could explain this tension.
%Adding tomographic bins to the analysis increases the number of COSEBIs modes needed, which 
%results in a less accurate estimation of covariance matrices from a set of simulations. 
%We improve our covariance matrix estimation by compressing COSEBIs 
%from 147 data points to 20 compressed COSEBIs and show the first measurement of CCOSEBIs from data. 
%We find significant B-modes in CFHTLenS data, particularly for large scales with tomography. 

\end{abstract}

% Select between one and six entries from the list of approved keywords.
% Don't make up new ones.

%%%%%%%%%%%%%%%%%%%%%%%%%%%%%%%%%%%%%%%%%%%%%%%%%%

%%%%%%%%%%%%%%%%% BODY OF PAPER %%%%%%%%%%%%%%%%%%

\section{Introduction}

Observations of weak gravitational lensing by the large-scale
structure in the Universe provides a powerful probe of dark matter, dark energy and modified
gravity theories. The underlying physics of lensing is well
understood, leaving the non-trivial measurement itself as the main
challenge in reaching the full potential of this cosmological tool.
Three major new weak lensing surveys are under way,
with the Kilo-Degree Survey (KiDS), the Dark Energy Survey (DES), and the
Hyper-Suprime Camera Survey (HSC). 
KiDS and DES recently presented their first ‘cosmic shear’ measurements \citep{Kuijken15,Becker15}.
These new surveys already cover several hundreds
of square degrees, but for now they still lack statistical precision
in comparison to their deeper but smaller area predecessor,
the Canada-France-Hawaii Telescope Lensing Survey, CFHTLenS
\citep{Heymans12}. As such this survey still provides the tightest
cosmological constraints from weak gravitational lensing.

The tension between the results of the CFHTLenS tomographic
analysis \citep{Heymans13} and the cosmological measurements
from the cosmic microwave background \citep{Planck15}
has been widely reported. It has been interpreted
in different ways as a sign for new physics 
\citep[see for example][]{Dossett15,Planck15MG_DE,BattyeMoss14}, 
the combined effects of baryonic feedback and neutrinos 
\citep[][]{HarnoisVanWaerbeke15,Kohlinger15},
or previously unknown systematic errors 
\citep[see for example][]{Spergel15,Verde13,Raveri15,Addison15}. 
In this paper we address the question of systematic errors by subjecting
the CFHTLenS data to a rigorous test for shear systematics using
“Complete Orthogonal Sets of E/B-mode Integrals” also known as
“COSEBIs”. Gravitational lensing can only produce E-modes and
any detected B-modes are due to either systematic errors or other
physical effects.\footnote{Whereas source clustering and lens-lens
coupling can in principle generate B-modes from lensing \citep{Schneider02,Hilbert09}, 
their amplitude is too small to be
significantly detected in current and future surveys.}

The formalism for COSEBIs was developed in \cite{SEK10}. 
COSEBIs provide a complete  set of functions
for efficiently separating E-modes from B-modes and hence allow for robust
systematics tests using the B-modes and a fairly compressed data
set. 
\cite{SEK10} and \cite{Eifler11} showed that a small
number of COSEBIs modes are enough to essentially capture the full
cosmological information using numerical analysis and mock data,
respectively. 
% \cite{Kilbinger13} and \cite{} applied the method of COSEBIs
% for a non-tomographic cosmic shear analysis in the CFHTLenS.
\cite{Asgari12} extended the method to tomographic
bins and showed that although a small number of COSEBIs
modes is enough for each redshift bin pair, in the presence of
many redshift bins the total number of COSEBIs needed is relatively
high. This is also true for all the other conventionally used
cosmic shear observables such as the two-point correlation functions
or the convergence power spectrum.

The most common approach to estimate covariance matrices is to use
mock data from numerical simulations, but the precision with which
this can be measured decreases with the number of observables 
\citep{Hartlap07,TaylorJoachimi14,SellentinHeavens15}. 
The requirement to minimize the number of observables prompted \cite{AS2014}  
to develop a compression method which reduces this
number substantially, without significant loss of information. In this
paper we show the first measurement of these compressed
COSEBIs, which are called CCOSEBIs. We also present the first measurement
of tomographic COSEBIs.

% The formalism for COSEBIs was developed in \cite{SEK10}. 

CFHTLenS is a 3D weak lensing survey, analysing $u^*$, $g'$, $r'$, $i'$, $z'$ multi-band data 
spanning 154 deg$^2$ from the CFHT Legacy Survey Wide Programme.  
Observed in sub-arcsecond seeing conditions, this survey was optimised for weak lensing science.   
Pixel-level data processing used the lensing-quality {\sc THELI} data reduction package \citep{Erben13}.  
PSF Gaussianised photometry provided precise photometric redshift distributions \citep{Hildebrandt12} with a reasonable
level of accuracy as scrutinised in \cite{Choi15} using a spectroscopic galaxy cross-correlation clustering analysis.
Weak lensing shear measurements were derived and calibrated
using the \emph{lens}fit Bayesian model-fitting method \citep{Miller13}.
A series of detailed systematics analyses were applied to the full data set, 
resulting in the rejection of a quarter of the survey 
area in order to satisfy strict systematic criteria \citep{Heymans12}.  

A number of different cosmological analyses have been carried out using CFHTLenS.
\cite{Kilbinger13} performed a two-dimensional analysis of the data using several cosmic shear
estimators, including COSEBIs, the statistic that forms the focus of this work.   
This 2D analysis was extended by
\cite{Fu14} who used COSEBIs in conjunction with the third order aperture mass statistic
to constrain cosmological parameters. Aside from the analysis of CFHTLenS,
\cite{Huff14} applied COSEBIs on Sloan Digital Sky Survey (SDSS) 
data to constrain $\sigma_8$ and $\Omega_{\rm m}h^2\;.$

Analyses of CFHTLenS that incorporated the redshift-dependence of the 
weak lensing signal started with a two-bin tomographic analysis in
\cite{Benjamin13} and  \cite{Simpson13}.  
This was followed by a finer six-bin tomographic analysis in \cite{Heymans13},  where the data was
modelled as a combination of a cosmological signal and a contaminating signal from the presence of intrinsic
galaxy alignments \citep[see also][for re-analyses of this data set]{MacCrann15,DES15_CP,Joudaki16}.  
These statistical analyses were based on measurements of the two-point shear correlation functions (2PCFs). 
Using only blue galaxies, for which the intrinsic alignment contamination is expected to be negligible, 
\cite{Kitching14} carried out a full 3-D power spectrum analysis of the survey.  
This power spectrum analysis was restricted to relatively large physical scales to minimise the effects of
baryon feedback on the non-linear matter power spectrum 
\citep[see][for example]{Semboloni13,Mead15,HarnoisVanWaerbeke15}.   
As shown in \cite{Kilbinger_review} there is excellent consistency between the different cosmological 
constraints derived by these varied statistical analyses of the CFHTLenS survey.  
The most stringent one, and also the most in tension with the CMB results 
is the 6-bin tomographic analysis of \cite{Heymans13}. 
We therefore focus our systematics analysis on this tomographic data set.

This paper is structured as follows: \sect\ref{sec:methods}
outlines the statistical methods, COSEBIs and CCOSEBIs, that are used in this analysis. 
\sect\ref{sec:results} contains the main results, where we show the measured COSEBIs  and CCOSEBIs. 
We quantify the measured B-modes using a $\chi^2$ analysis 
and finally conclude in \sect\ref{sec:conclusions}. 
We verify our pipeline tests on mock data in the Appendix.

\section{Methods: COSEBIs and CCOSEBIs}
\label{sec:methods}

Converting a measured gravitational lensing shear field to a convergence field does not necessarily result in the 
real projected mass field expected from gravitational lensing theory
\citep[see][for a review of weak gravitational lensing]{BartelmannSchneider01}.
The reason is that aside from first order lensing effects there are other influential factors. 
These other factors fall into two categories according 
to whether their origin is physical or non-physical. 
The former may arise from higher-order lensing effects 
(contributions beyond the Born approximation, 
see \citealt{Shcneider98}), and source redshift clustering (\citealt{Schneider02}), 
or intrinsic galaxy alignments \citep[see][and references therein]{Blazek11}; 
The latter case involves noise contributions and remaining systematic effects, 
for example, in galaxy shape measurements. 
First order weak gravitational lensing can only produce 
modes which are commonly referred to as E-modes,
whereas, the modes which arise from the imaginary part of the estimated convergence field, 
$\kappa$, are called B-modes.
These modes are so named  because of the similar mathematical properties 
of the shear field and the polarization of an electromagnetic radiation field (both of them are polars). 
B-mode contributions from physical effects are expected to be negligible for a survey like CFHTLenS.
Hence any detection of a B-mode will arise from either inaccuracies in the shape measurements and/or selection biases.
Since the physical contributions to the B-modes are very small, measuring a statistically zero B-mode, 
suggests (but does not guarantee) a satisfactory PSF correction. Separating these modes 
is essential to test for systematic errors. 

Any observable (statistic) which separates E-modes from B-modes at the two-point statistics level, 
can be written in the following form,
\begin{align}
\label{eqE-B}
E = \frac{1}{2} \int_0^{\infty}\mathrm{d}\vartheta\:\vartheta\:
[T_+(\vartheta)\xi_+(\vartheta) + T_-(\vartheta)\xi_-(\vartheta)]\;,\\ \nonumber
B= \frac{1}{2} \int_0^{\infty}\mathrm{d}\vartheta\:\vartheta\:
[T_+(\vartheta)\xi_+(\vartheta) - T_-(\vartheta)\xi_-(\vartheta)]\;,
\end{align} 
where $\xi_{\pm}(\vartheta)$ are the two-point correlation functions (2PCFs) 
of the shear field, $\vartheta$ is the angular distance 
between pairs of galaxies on the sky and $ T_{\pm}(\vartheta)$ are filter functions, 
that are chosen to produce pure E/B-modes, corresponding to $E$/$B$, respectively. 
In \cite{SchneiderKilbinger07}, conditions for such filters were obtained,
\begin{flalign}
\label{T-cond}
\int_{\vartheta_{\rm{min}}}^{\vartheta_{\rm{max}}}
\frac{\mathrm{d}\vartheta}{\vartheta}\:T_-(\vartheta) & =0= 
\int_{\vartheta_{\rm{min}}}^{\vartheta_{\rm{max}}}
\frac{\mathrm{d}\vartheta}{\vartheta^3}\:T_-(\vartheta)\;,\\
\label{T+cond}
\int_{\vartheta_{\rm{min}}}^{\vartheta_{\rm{max}}}
\mathrm{d}\vartheta\:\vartheta\:T_+(\vartheta) & =0= 
\int_{\vartheta_{\rm{min}}}^{\vartheta_{\rm{max}}}
\mathrm{d}\vartheta\:\vartheta^3\:T_+(\vartheta)\;,
\end{flalign}
where $\vartheta_{\rm{min}}>0$ and $\vartheta_{\rm{max}}$ is finite. 
Using these conditions \cite{SEK10} constructed two complete orthogonal sets of 
filter functions, $T_{\pm}$ which form the basis of the COSEBIs.

\subsection{COSEBIs}

The two sets of COSEBIs basis functions are the Lin- and Log-COSEBIs, 
which are written in terms of polynomials in $\vartheta$ and $\ln(\vartheta)$ in real space, respectively.
In addition to \cite{SEK10}, \cite{Fu10} constructed filters which maximized the signal-to-noise ratio 
for a specific angular range, or maximized the information content of 
$E$ statistics via Fisher analysis. 
In this analysis we use the Log-COSEBIs, as they require fewer modes compared to the Lin-COSEBIs 
to essentially capture all the information 
(see \citealt{SEK10} for a single redshift bin and \citealt{Asgari12} for the tomographic case). 

The COSEBIs can be written in terms of the 2PCFs in real space,
\begin{align}
\label{eq:EnReal}
 E_n^{(ij)} &= \frac{1}{2} \int_{\theta_{\rm min}}^{\theta_{\rm max}}
 \d\vartheta\,\vartheta\: 
 [T_{+n}(\vartheta)\,\xi^{(ij)}_+(\vartheta) +
 T_{-n}(\vartheta)\,\xi^{(ij)}_-(\vartheta)]\;, \\
 \label{eq:BnReal}
 B_n^{(ij)} &= \frac{1}{2} \int_{\theta_{\rm min}}^{\theta_{\rm
     max}}\d\vartheta\,\vartheta\: 
 [T_{+n}(\vartheta)\,\xi^{(ij)}_+(\vartheta) -
 T_{-n}(\vartheta)\,\xi^{(ij)}_-(\vartheta)]\;,
\end{align} 
where $E_n^{(ij)}$ and $B_n^{(ij)}$ are the E and B-mode COSEBIs for redshift bins $i$ and $j$
, $T_{\pm n}(\vartheta)$ are the COSEBIs filter functions and $n$, a natural number, is the order of the COSEBIs modes.
The modes with larger $n$ values are typically more sensitive to
small-scale variations in the shear 2PCFs, while the modes with small $n$ are sensitive to large-scale variations.
This is because $T_{\pm n}$ are oscillatory functions with $n+1$ roots in their range of support.
Alternatively, the E/B-COSEBIs can be expressed as a function of the convergence power spectra:
\begin{align}
\label{eq:EnFourier}
E_n^{(ij)} &= \int_0^{\infty}
\frac{\d\ell\,\ell}{2\pi}P^{(ij)}_{\mathrm{E}}(\ell)W_n(\ell)\;,\\
\label{eq:BnFourier}
B_n^{(ij)} &= \int_0^{\infty}
\frac{\d\ell\,\ell}{2\pi}P^{(ij)}_{\mathrm{B}}(\ell)W_n(\ell)\;,
\end{align} 
where $P^{(ij)}_{\mathrm{E(B)}}$ are the E(B)-mode convergence power spectra and
the $W_n(\ell)$ are the Hankel transform of $T_{\pm n}(\vartheta)$ 
\begin{align}
\label{Wn}
W_n(\ell) & =  \int_{\vartheta_{\rm{min}}}^{\vartheta_{\rm{max}}}\d\vartheta\:
\vartheta\:T_{+n}(\vartheta) \rm{J}_0(\ell\vartheta) \nonumber \\ 
& = \int_{\vartheta_{\rm{min}}}^{\vartheta_{\rm{max}}}\d\vartheta\:
\vartheta\:T_{-n} (\vartheta) \rm{J}_4(\ell\vartheta)\;,
\end{align} 
with $\rm{J}_0$ and $\rm{J}_4$ as the ordinary Bessel functions of zeroth and fourth order.

We use \Eqt\eqref{eq:EnFourier} to find the theory value of the E-mode COSEBIs    
as most theories provide us with an input power spectrum. 
However, in practice the shear 2PCFs are more straightforward to measure from data, hence, 
\Eqt\eqref{eq:EnReal} and \Eqt\eqref{eq:BnReal} are used to calculate the E/B-mode COSEBIs from 
data and simulations.

\subsection{Compressed COSEBIs: CCOSEBIs}

Data compression is a challenge that will become increasingly more important for future large scale surveys such as 
Euclid\footnote{http://sci.esa.int/euclid/, \cite{EuclidRB11}} and LSST\footnote{http://www.lsst.org/lsst/}.
The main reason data compression is essential is that
the number of simulations needed to estimate the data covariance matrix accurately, depends on 
the number of observables. Therefore, having a smaller set of observables reduces the number of 
cosmological simulations needed. 

\cite{AS2014} developed a compression method which is based on the sensitivity 
of observables (statistics) to the parameters to be measured. 
This method relies on our understanding of these parameters, since the compressed observables
depend  on the covariance and derivatives of the parent observable to the parameters at their fiducial value. 
The assumption behind this compression method is that we have a relatively good idea of the value 
of the parameters that we want to measure (for example from previous observations), 
which is correct for most of the cosmological parameters. 
One might expect to lose a significant portion of the information about 
the parameters if the fiducial covariance matrix used for constructing the parameters is not close to the truth. 
However, \cite{AS2014} applied this compression method to tomographic COSEBIs and
showed that the weak lensing information lost due this compression is small even for very inaccurate 
COSEBIs covariance matrices. 
This implies that this compression is insensitive to the inaccuracies in the estimated
covariance matrix of the parent observables, 
which means that using this compression allows for the same accuracy
in estimations with fewer cosmological simulations.

Here we will also use compressed COSEBIs (CCOSEBIs) for the analysis of the CFHTLenS data. 
The CCOSEBIs are linear combinations of the COSEBIs. 
The coefficients of these linear combinations are written in terms of the covariance and the
derivatives of the COSEBIs with respect to cosmological parameters,
\begin{equation}
\label{eq:CCOSEBIs}
 \boldsymbol{E}^{\rm c}= \boldsymbol{\Gamma} \boldsymbol{E}\;,
\end{equation} 
where $\boldsymbol{E}^{\rm c}$ is the E-mode CCOSEBIs vector, 
$\boldsymbol{E}$ is the $E_n^{ij}$ vector 
and $\boldsymbol{\Gamma}$ is the compression matrix defined as,
\begin{equation}
\label{eq:CompressionMatrix}
 \boldsymbol{\Gamma}\equiv\HMb \CMb^{-1}\;,
\end{equation}
where $\HMb$ is a matrix formed of both first and second derivatives of the
COSEBIs with respect to the cosmological parameters and $\CMb$ is 
the covariance matrix of COSEBIs \citep[see section 2 of][for the details of the formalism]{AS2014}. 
The number of CCOSEBIs modes for constraining $P$ cosmological parameters is $P(P+3)/2$, 
regardless of the number of COSEBIs used. 
For a total of $\N_{\rm max}$ COSEBIs modes and $P$ parameters, 
$\boldsymbol{\Gamma}$ is a matrix with $P(P+3)/2$ 
rows and $\N_{\rm max}$ columns, where the first $P$ rows are the COSEBIs first order derivatives 
while the last $P(P+1)/2$ rows are the second order derivative of COSEBIs with
respect to the parameters.

%{\it Side note for authors: COSEBIs have all the information about the B-modes that can be separated. 
%Theoretically we need to use all of the $B_n$ (n=1,2 .... ) to find all of it. 
%For the E-modes we know that a small number would suffice but we don't know that for the B-modes. 
%That aside the CCOSEBIs are linear combinations of the COSEBIs 
%and the coefficients of this linear combination depends on the covariance and the derivatives 
%of E-mode COSEBIs to the cosmological parameters. 
%These coefficients are designed to find E-modes efficiently, but not B-modes. 
%B-modes have no effect on the coefficients. 
%I think it is not that surprising that they can see the B-modes so well. 
%However, they also show better fits to the E-modes compared to COSEBIs.}

\section{Results}
\label{sec:results}

In this section we apply two analysis methods, based on COSEBIs and CCOSEBIs respectively, to measure the cosmic shear signal from CFHTLenS data. 
Before applying our methods on the data we performed a number of tests including blind tests on mock data, 
as explained in the Appendix. The Appendix also details the technical aspects of calculating the COSEBIs from 
shear two-point correlation functions. 

\subsection{Analysis}
\label{sec:AnalysisSetups}

In order to compare our results with the previous CFHTLenS analysis
as well as to test the data for systematic errors in a comprehensive manner,
we analyse the data in several different ways.
We choose the three angular ranges, $[1',40']$, $[40', 100']$ and $[1', 100']$ corresponding to
 small, large and the combination of both angular scales.  
We also consider two sets of galaxy populations, all and blue galaxies only. 
The blue galaxies are late-type galaxies and are expected to have 
a negligible intrinsic galaxy alignment signal \citep[see][]{Heymans13}.
This population is selected using their Bayesian photometric redshift spectral type, $T_B>2$ \citep[see][for the definition]{Velander14}. 
In addition, we compare a 2D, non-tomographic, analysis with a 6 redshift bins tomographic analysis.
\tab\ref{tab:CFHTLenSNeff} shows the redshift bins and their corresponding effective number density of galaxies
for the blue and all galaxies. The redshift distribution of the CFHTLenS data 
is measured using photometric redshift estimates as explained in \cite{Hildebrandt12}.

% following previous CFHTLenS analyses, but refer the reader to ,
% where errors are determined and shown to be non-negligible.

% Consistency between the results for different angular ranges,
% is a great way to assure that analysis is reliable.

\begin{table}
\centering
 \caption{\small{Effective number density of galaxies, $n_{\rm eff}$ (arcmin$^{-2}$), 
 in each redshift bin for late-type (Blue) and all (All) Galaxies.}}
 \label{tab:CFHTLenSNeff}
 \begin{adjustbox}{max width=\textwidth}
 \begin{tabular}{  c  c c }
  \hline
  z-bin        & Blue: $n_{\rm eff}$   & All: $n_{\rm eff}$ 	\\ \hline
  [0.2, 0.39]  & 1.507                & 1.811           	\\ \hline
  [0.39, 0.58] & 1.265                & 1.646           	\\ \hline
  [0.58, 0.72] & 1.560                & 1.907           	\\ \hline
  [0.72, 0.86] & 1.366                & 1.788           	\\ \hline
  [0.86, 1.02] & 1.440                & 1.729           	\\ \hline
  [1.02, 1.3]  & 1.395                & 1.708           	\\ \hline
  [0.2,  1.3]  & 8.533                & 10.589          	\\ \hline
 \end{tabular}
 \end{adjustbox}
\end{table}

Listed below are the configurations we used in this paper which best resemble the
previous two-point statistics cosmic shear analysis of CFHTLenS. 
\begin{itemize}
\item \cite{Heymans13} performed an analysis with a set-up, which
corresponds to the tomographic $[1',40']$ angular range with all galaxies. 
They modelled galaxy intrinsic alignments with a single parameter, as the intrinsic-shear 
signal is non-negligible when all galaxies are considered in tomographic bins. 
 
\item \cite{Kitching14} used large scales (roughly the $[40', 100']$ range) with blue galaxies. 
They used 3D cosmic shear analysis in Fourier space which is approximately equivalent to our tomographic analysis. 

\item \cite{Kilbinger13} used a large range of scales for their analysis which is close to the $[1', 100']$ range we consider. 
Their analysis considered all galaxies without any redshift binning. 

\end{itemize}

{  In this analysis we choose to ignore the CFHTLenS photometric redshift biases and uncertainties presented in \cite{Choi15}
in order to be able to directly compare our results to the CFHTLenS analyses listed above. 
\cite{Joudaki16} investigated the effect of the redshift biases and showed that the effect is small on the cosmological information. 
The B-mode analysis, which is the main subject of this work, is essentially unaffected by the redshift measurement biases. }

\subsection{Cosmological models}

The cosmological models we compare our results to are two flat $\Lambda$CDM models, with parameters corresponding to the 
best fit values of CFHTLenS+WMAP7 \citep[]{Heymans13} and Planck TT+ lowP \citep[]{Planck15}.
We assume a primordial power-law power spectrum and use the \cite{BondEfstathiou84} transfer function
to calculate the linear matter power spectrum. The non-linear power spectrum is estimated 
using the halo fit formula of \cite{Smith03}.  \cite{MacCrann15} show that this choice of non-linear fitting function does not
significantly change cosmological parameter constraints with CFHTLenS, in comparison to analyses that use improved
non-linear correction schemes \citep{Takahashi12, Mead15}.  

The cosmological parameters are given in \tab\ref{tab:CosmoParam},
where we also show the parameters for the simulation products which are used for pipeline verifications as well as estimating
the covariances. The cosmological parameters which are presented in \tab\ref{tab:CosmoParam} are, $\sigma_8$, the normalization
of the matter power spectrum, $\Omega_{\rm m}$, the mean matter density parameter, $n_{\rm s}$,
the spectral index, $h$, the dimensionless Hubble parameter and $\Omega_\mathrm{b}$, the baryonic matter density 
parameter. Spatial flatness is assumed throughout out this work, which means that $\Omega_{\Lambda}=1- \Omega_{\rm m}$, where 
$\Omega_{\Lambda}$ is the dark energy density parameter. 

\begin{table}
\centering
\caption{\small{Cosmological parameters for a flat $\Lambda$CDM cosmology. 
The first row corresponds to CFHTLenS+WMAP7 best fit values, 
the second row belongs to Planck best fit values for TT+lowP and 
the final row shows the values for the SLICS simulations.}}
\label{tab:CosmoParam}
\begin{tabular}{ c  c  c  c  c  c  c  }
%   \cline{2-7}
             & $\sigma_8$       & $\Omega_{\rm m}$        & $n_\mathrm{s}$         & $h$               & $\Omega_\mathrm{b}$ \\
  \hline
  CF+WM      & 0.794            &  0.255                  &  0.967                 & 0.717             & 0.0437 \\
  \hline
  Planck     & $0.829$          &  $0.315$                &  $0.9655$              & $0.6731$          & $0.0490$ \\
  \hline
  SLICS      & 0.826            &  0.2905                 &  0.969                 & 0.6898            & 0.0473 \\
  \hline
\end{tabular}
\end{table}

\subsection{Covariance}

The covariance matrix of the COSEBIs is measured from mock galaxy 
catalogues constructed from the SLICS, a suite of N-body simulations described in 
\cite{Harnois15}. The mock galaxy population algorithm, 
detailed in \cite{Joudaki16}, is designed to reproduce the properties of the CFHTLenS catalogues. 
These new mock catalogues are updated versions of those used in the previous analysis of the CFHTLenS, 
which offer better precision especially at large angular scales, since the box size of the simulations is L = 505 Mpc/h; 
which is significantly larger than the simulation 
set used for modelling the earlier CFHTLenS measurements (L = (147,231) Mpc/h),
hence, the new simulation set is less affected 
by suppression of the large-scale variance by finite box size effects. 
Furthermore, we use 497 in comparison to the 184 independent simulations used in the earlier work. 

Estimating covariances from a finite number of 
simulations is noisy which causes biases in the inverse covariance \citep[see][]{Hartlap07}.
Assuming Gaussian errors on the estimated covariance matrix, $\tens{\hat C}$, 
the inverse covariance matrix is given by
\begin{equation}
\label{eqCorrection}
 \tens{C}^{-1}=\frac{n_{\rm sim}-n_{\rm obs}-2}{n_{\rm sim}-1}\tens{\hat C}^{-1}\;,
\end{equation}
where $n_{\rm sim}$ and $n_{\rm obs}$ are the number of simulations and observables, respectively.  
For $n_{\rm obs}/n_{\rm sim}<0.8$, the above formula produces an unbiased inverse covariance 
according to \cite{Hartlap07}.  It will however still have noise associated with it, which depends on the ratio
of the number of observables to the number of simulations. 
\cite{TaylorJoachimi14} extended this analysis by providing a more accurate correction for the parameter covariance
matrix as,
\begin{equation}
\label{eqCorrectionParameter}
 \tens{C}_{\rm par}=\frac{n_{\rm sim}-n_{\rm obs}-2}{n_{\rm sim}-n_{\rm obs}+n_{\rm par}-1}
 \tens{\hat C}_{\rm par}\;,
\end{equation}
where $\tens{C}_{\rm par}$ is the parameter covariance matrix and $n_{\rm par}$ is the number of 
parameters to be estimated. Applying this correction to $\tens{C}_{\rm par}$ results in a slightly
smaller covariance matrix in comparison to the \cite{Hartlap07} method, for $n_{\rm par}<<(n_{\rm sim}-n_{\rm obs})$, 
but there is still noise associated with it.
\cite{SellentinHeavens15} extended this analysis further to mitigate covariance matrix estimation uncertainties 
by marginalising over the true covariance matrix given its estimated value. 
They show improvements over the \cite{Hartlap07} and \cite{TaylorJoachimi14}  estimate, by 
noting that their corrected covariance matrix distribution is no longer Gaussian.

In our analysis, the maximum number of observables that we use
is $7\times21=147$ COSEBIs modes, where $7$ is the number of COSEBIs modes in each redshift pair 
and $21$ is the number of redshift pairs for the tomographic case.
As a result the ratio $n_{\rm obs}/n_{\rm sim}\approx0.3$, 
which can cause about $7\%$ errors in the estimated
inverse covariance using the \cite{Hartlap07} correction. 
{ This value for the error on the covariance matrix is acceptable for analysing
CFHTLenS data around the maximum likelihood point. 
However, around the tails of the likelihood distribution the \cite{SellentinHeavens15} correction becomes significant.
Therefore, we apply this correction in \sect\ref{sec:FoM}, where we calculate the p-values, 
primarily to assess the significance of the detected B-modes.}
% so we do not implement the further corrections developed by \cite{SellentinHeavens15} and \cite{TaylorJoachimi14}. 
% For precision cosmology, however, these errors will not 
% acceptable (see \citealt{EuclidRequirements10} for Euclid mission requirements), 
% further motivating data compression with CCOSEBIs.

\subsection{Measurements}

Following \cite{Heymans12}, we analyse  the 129 CFHTLenS fields
that passed the systematic tests, representing 75\% of the total observed area.

We calibrate the data correcting for additive and multiplicative biases between the observed, $\epsilon_{\rm obs}$, 
and the true ellipticities, $\epsilon_{\rm true}$, modelled as
\begin{equation}
 \epsilon_{\rm obs}=(1+m)\:\epsilon_{\rm true}+ c\;,
\end{equation}
where $\epsilon$ is a complex qunatity defined as $\epsilon=\epsilon_1+i\epsilon_2$, 
where $\epsilon_1$ and $\epsilon_2$ are real quantities.

In CFHTLenS analyses, $c$ was measured to be zero and 
$~2\times10^{-3}$ on average for $\epsilon_1$ and $\epsilon_2$ respectively.
The origin of the additive bias is unknown and its value is calibrated from the data empirically. 
It is likely that the multiplicative bias, $m$, 
originates from the effect of noise in shape measurements \citep[see for example][]{Melchior_Viola_2012}. 
It is estimated from galaxy image simulations.
While the additive bias is subtracted from the observed $\epsilon_2$ directly, the effect
of the multiplicative bias is applied globally as explained in \cite{Miller13}. 
The measured 2PCFs are divided by the calibration function,
\begin{equation}
 1+K(\vartheta)=\frac{\sum_{ab} w_a w_b(1+m_a)(1+m_b)}{\sum_{ab} w_a w_b}\;,
\end{equation}
where $w_a$ and $m_a$ are the weight and the multiplicative bias associated with a galaxy 
at position $a$. The sum is carried out over all pairs of galaxies with a separation falling within the $\vartheta$ bin.
Each galaxy has an inverse variance weight associated with it. 
Less noisy galaxy shapes have a larger weight value, ergo they are more important
in the analysis. The definition of $w$ can be found in \cite{Miller13}. 

The estimated 2PCFs, from the input ellipticities and 
their associated weights, $w$, for redshift bins $i$ and $j$, are given by
\begin{equation}
 \hat\xi_\pm^{ij}(\vartheta)=\frac{\sum w_a w_b
 \left[\epsilon_{\rm t}^i({\pmb x}_a)\epsilon_{\rm t}^i({\pmb x}_b)
 \pm\epsilon_{\rm x}^i({\pmb x}_a)\epsilon_{\rm x}^i({\pmb x}_b)\right]}
 {\sum w_a w_b}\;,
\end{equation}
where $\epsilon_{\rm t/x}({\pmb x}_a)$ are the tangential/cross 
ellipticities at position ${\pmb x}_a$,
with respect to the reference frame connecting the pairs of galaxies involved.
% The sum is taken over all pairs of galaxies with separation within a bin around $\vartheta$.
$\hat\xi_\pm^{ij}(\vartheta)$ is then divided by $1+K(\vartheta)$ to find an unbiased estimate.

To estimate the 2PCFs, we use Athena\footnote{http://www.cosmostat.org/software/athena/}\citep[see][]{KilbingerAthena14}
a tree code that calculates second-order correlation functions from input galaxy  catalogues. 
The opening angle that we use is 0.02 radians, which shows no significant differences 
with a brute force (opening angle=0) estimation.

%The algorithm for the tree code in this case divides the area of a field of galaxies 
%into cells of different sizes.
%If two galaxies are in two different cells with a large distance
%then they are assumed to be in the centre of the cell. This method enables a faster
%2PCFs calculation, at the loss of accuracy. However, the algorithm accepts an input parameter
%called the opening angle which controls the cell size. 
%The smaller the opening angle the more accurate and slower the calculations. 
%Here I use an opening angle of 0.02 radians. 
% We have also compared the results with a brute force calculation (opening angle = 0)  and 
% found no significant differences. }

The estimated 2PCFs are then inserted into 
Eqs.\thinspace\eqref{eq:EnReal} and \eqref{eq:BnReal} to determine the COSEBIs 
E and B-modes, respectively (the details of which are explained in the Appendix). 
The theory values of COSEBIs are estimated using Eqs.\thinspace\eqref{eq:EnFourier} 
and \eqref{eq:BnFourier} which relate the COSEBIs to the convergence power spectrum directly.
In this analysis we use the first 7 COSEBIs modes, since \cite{Asgari12} have shown that 
these are enough to essentially capture the full information for up to 
7 cosmological parameters\footnote{  Depending on the origin of the B-mode systematic, 
7 COSEBIs modes may not be enough to capture all of the information in the B-mode signal.  
Further work is required to test different systematic scenarios and how they impact the different COSEBIs. 
For the purpose of this paper, however, we match our B-mode analysis to the 7 modes 
that are optimal for E-mode measurements.}.
Assuming tomographic bins each redshift bin pair will have 7 COSEBIs modes which adds up to 
147 modes in total. Using the compression method in \cite{AS2014} we decrease this number to 20.  

\begin{figure}
  \begin{center}
    \begin{tabular}{c}
      \resizebox{80mm}{!}{\includegraphics{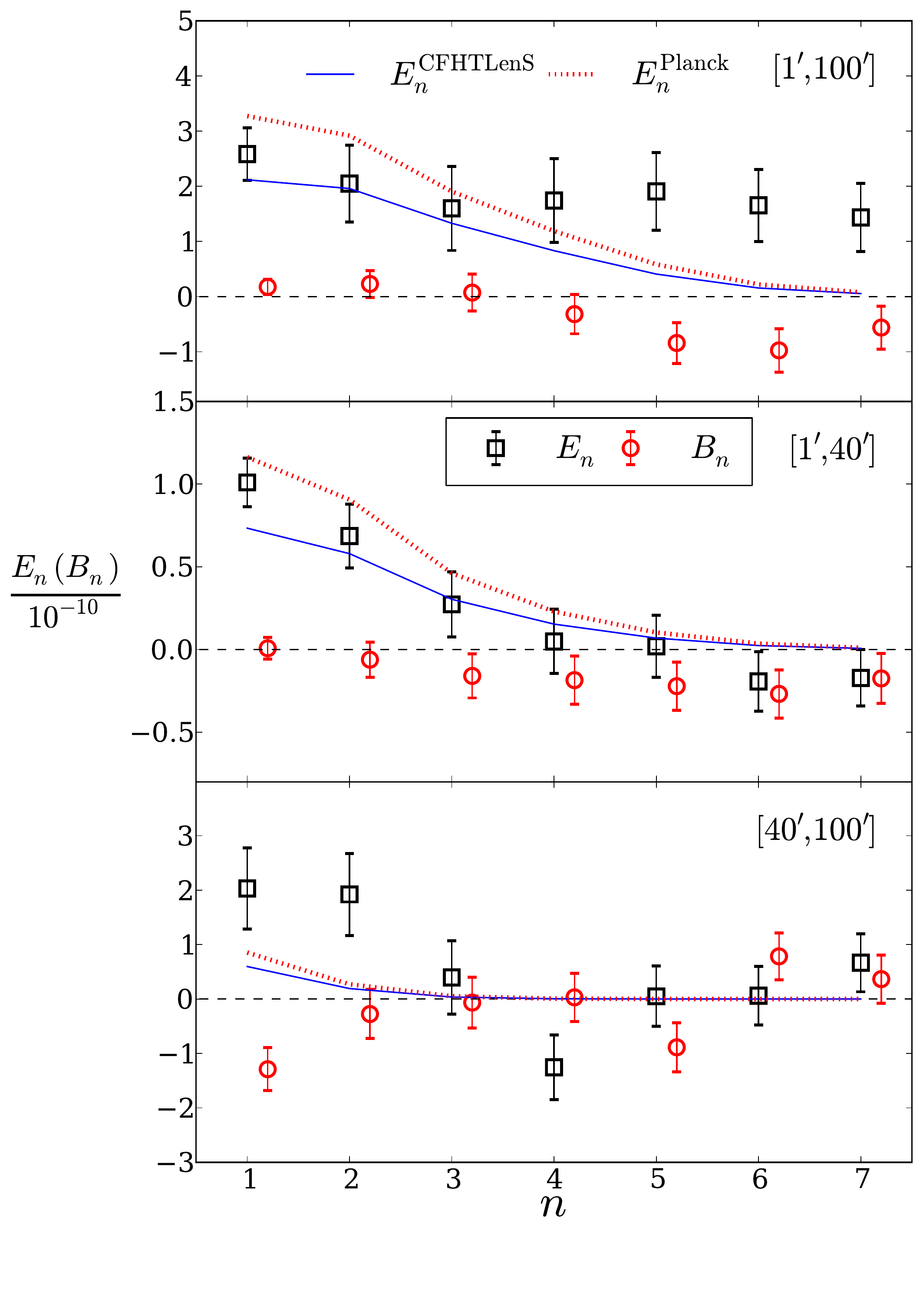}}
    \end{tabular}
    \caption{\small{Measured COSEBIs from the CFHTLenS data for a single redshift bin using all galaxies. 
    Three angular ranges are considered here. The dashed line shows the zero B-mode value. 
    The $B_n$ modes (red circles) are shifted to the right for visual assistance. 
    The $E_n$ (black squares) are compared with their theoretical values given the Planck (red dotted curve)
    and CFHTLenS+WMAP7 (blue solid curve) cosmologies. {  The CFHTLenS+WMAP7 theoretical values are best fit values
    for the $[1', 40']$ angular range with tomography \citep[see][]{Heymans13}.}
    The values of the cosmological parameters for the theoretical curves are given in
    \tab\ref{tab:CosmoParam}. 
    Note that the COSEBIs modes are discrete and the theory values are connected to each other for visual inspection.
    The errors are estimated from simulated data explained in \App\ref{app:SLICS}. 
    Note that the different modes are correlated (see the covariance in \fig\ref{fig:CovSim1bin})}}
    \label{fig:CFHTLenS1bin}
  \end{center}
\end{figure}

\fig\ref{fig:CFHTLenS1bin} shows the measured COSEBIs for a single redshift bin using all galaxies.  
The panels show the results for the three angular ranges, $[1',100']$, $[1', 40']$ and $[40', 100']$. 
The symbols show the COSEBIs modes estimated from the data while the theory values are shown as curves. 
The COSEBIs modes are discrete and the curves are drawn to aid the viewer. The E-mode COSEBIs are shown
by black squares while the red circles are the B-modes. The B-modes are shifted to the right to aid the viewer. 
The errors on the data are estimated from the simulations and are correlated (see the covariance in \fig\ref{fig:CovSim1bin}). 
As we will see in \sect\ref{sec:FoM}, the B-modes in this plot are only significant for the angular scale  $[40', 100']$.
The theory E-mode curves belong to CFHTLenS+WMAP7 and Planck best fit values listed in \tab\ref{tab:CosmoParam}.
We also see that the highest signal-to-noise ratio comes from small scales as expected \citep[see][for example]{Asgari12}.

\begin{figure*}
  \begin{center}
    \begin{tabular}{c}
      \resizebox{175mm}{!}{\includegraphics{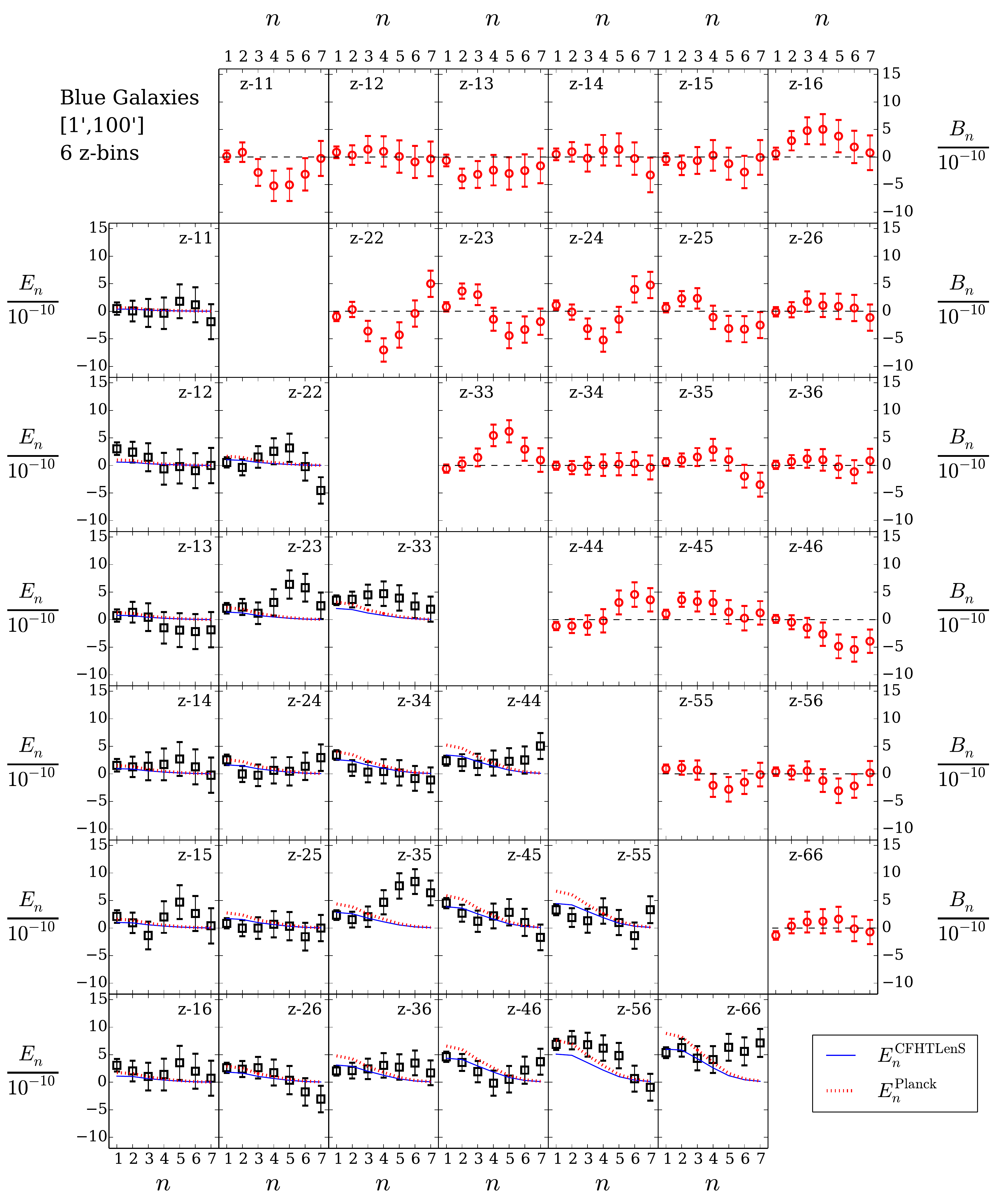}}
    \end{tabular}
    \caption{\small{Measured COSEBIs from the CFHTLenS data for 6 redshift bins using blue galaxies. 
    The angular range $[1', 100']$ is used here. The B-modes (red circles) are shown in the upper right triangle,
    while the E-modes (black squares) are shown in the lower left triangle for the redshift bin pairs
    indicated for each panel. The theoretical values of $E_n$ are shown for the Planck (red dotted curve)
    and CFHTLenS+WMAP7 (blue solid curve) cosmologies (see \tab\ref{tab:CosmoParam}). 
    Note that the COSEBIs modes are discrete and the theory values are connected to each other for 
    visual inspection.
    The errors are estimated from the mock data explained in \App\ref{app:SLICS}. 
    Note that the different modes are correlated as shown in \fig\ref{fig:CovSim6bin}.}}
    \label{fig:CFHTLenS6Bins}
  \end{center}
\end{figure*}

\fig\ref{fig:CFHTLenS6Bins} shows the estimated COSEBIs for the tomographic case with blue galaxies. 
The E/B-modes are separated into the upper and lower triangle of the plot. Each panel belongs to a redshift bin pair
indicated at its corner. 
Similar to \fig\ref{fig:CFHTLenS1bin} the measured E and B-modes are shown as black squares and
red circles, respectively. 
The curves show the theory values of the E-modes for the CFHTLenS+WMAP7 and Planck cosmologies in 
\tab\ref{tab:CosmoParam}. The angular range considered is $[1', 100']$.
Unlike the single redshift bin case, we see statistically non-zero B-modes in this Figure.

\begin{figure*}
  \begin{center}
    \begin{tabular}{c}
      \resizebox{175mm}{!}{\includegraphics{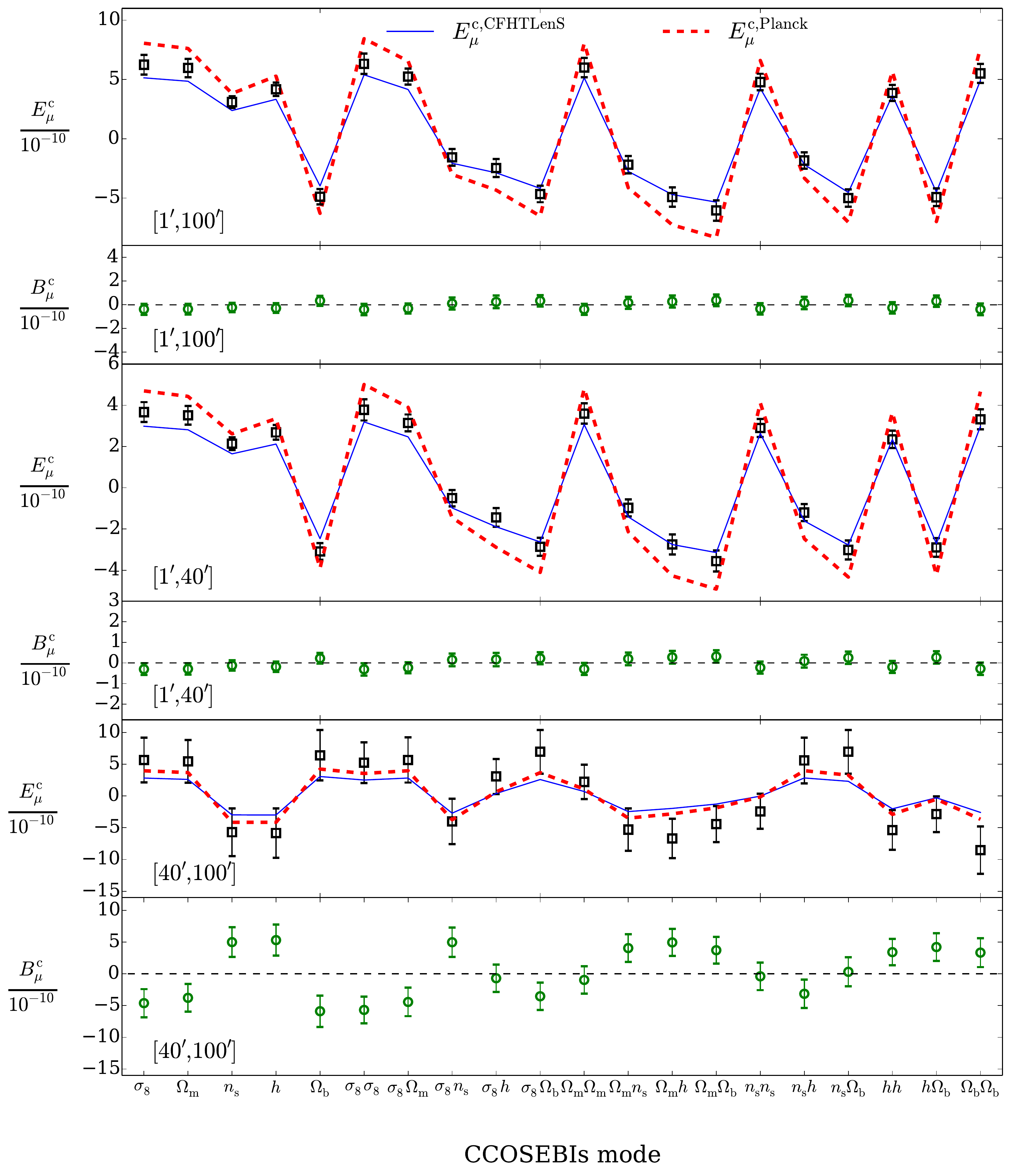}}
    \end{tabular}
    \caption{\small{Measured CCOSEBIs from the CFHTLenS data for 6 redshift bins using blue galaxies. 
    The B-modes are shown as green circles.
    The black dashed line shows where the zero line for the B-modes lies.
    The measured E-modes are shown as black squares, 
    while the theory values corresponding to the best fit values for  CFHTLenS+WMAP7
    and Planck (see \tab\ref{tab:CosmoParam}) cosmologies are shown as blue solid curves and red dotted curves, respectively.    
    Note that the CCOSEBIs modes are discrete and the theory values are connected to each other for visual inspection.
    The errors are estimated from simulated data explained in \App\ref{app:SLICS}. 
    Note that the different modes are correlated (see the covariance in \fig\ref{fig:CovSimCCOSEBIs}).}}
    \label{fig:CFHTLenSCCOSEBIs}
  \end{center}
\end{figure*}

\fig\ref{fig:CFHTLenSCCOSEBIs} shows the first measurement of CCOSEBIs from data. 
We use blue galaxies with 6 tomographic bins and the three angular ranges to estimate the CCOSEBIs. 
Here we choose the 5 cosmological parameters in \tab\ref{tab:CosmoParam}
to compress COSEBIs into 5 first order and 15 second order CCOSEBIs, using the Planck values as our
fiducial cosmology to calculate the compression matrix (see \Eqt\ref{eq:CompressionMatrix}). 
The CCOSEBIs modes are named after the parameters which are used to define them, shown on the x-axis. 
The first order modes only depend on one cosmological parameter, whereas, 
the second order CCOSEBIs depend on two parameters which could be the same. 
For example, the points related to $\Omega_{\rm m} h$ show the value of the second order CCOSEBIs mode which 
is based on the derivatives of COSEBIs to $\Omega_{\rm m}$ and  $h$. 
The ordering of the modes is arbitrary and the apparent oscillations in the figure can be rearranged.
The theory values of the CCOSEBIs for CFHTLenS+WMAP7 
and Planck cosmologies are shown as the blue solid curve and the red dashed curve.
Note that the CCOSEBIs modes are discrete and the theory values are connected for an easier comparison. 
The B-modes are shown on the same scale as the E-modes.
The CCOSEBIs are designed to be sensitive to cosmological information about these parameters. 
Therefore, they may not be as sensitive to the B-modes in the data. 
As we will see for most cases that we have studied, even if there are significant B-modes picked up by COSEBIs, 
the B-modes are not always significant with the CCOSEBIs. 
The exception is the $[40',100']$ angular range which shows significant B-modes either way.

\subsection{Figure-of-merit and fitting}
\label{sec:FoM}
To quantify the significance of the measured B-modes we estimate their $\chi^2$ value with zero,
\begin{equation}
 \chi^2_{B}=\sum\boldsymbol{B}^{\rm t} \CMb^{-1} \boldsymbol{B}\;,
\end{equation}
where $\boldsymbol{B}$ is a vector composed of $B_n$, $\boldsymbol{B}^{\rm t}$ is its transpose and 
$\CMb^{-1}$ is the inverse of the B-mode covariance matrix, estimated from the SLICS simulations. 
We also estimate the $\chi^2$ values for the E-modes compared to the best fit values of CFHTLenS+WMAP7
and Planck (see \tab\ref{tab:CosmoParam}). The raw value of the $\chi^2$ is not particularly informative, even when the
degrees-of-freedom is known \citep[see][for example]{Andrea10}. Hence instead we show the p-values for the estimated $\chi^2$ values.
The p-value shows the probability of finding a $\chi^2$ value larger than the one estimated. We choose a significance
level of $99\%$, p-value=0.01, which corresponds to a deviation of about $2.6\sigma$ for a normal distribution. Recall that
a $\chi^2$ distribution is skewed towards smaller values and asymptotically reaches a normal distribution for 
large numbers of degrees-of-freedom as illustrated in \fig\ref{fig:ChiSDist}. 
{  Additionally, using an inverted noisy covariance changes a $\chi^2$ distribution and hence  the derived p-values, which
  we account for using the method proposed by \cite{SellentinHeavens15}}.

\begin{figure}
  \begin{center}
    \begin{tabular}{c}
      \resizebox{80mm}{!}{\includegraphics{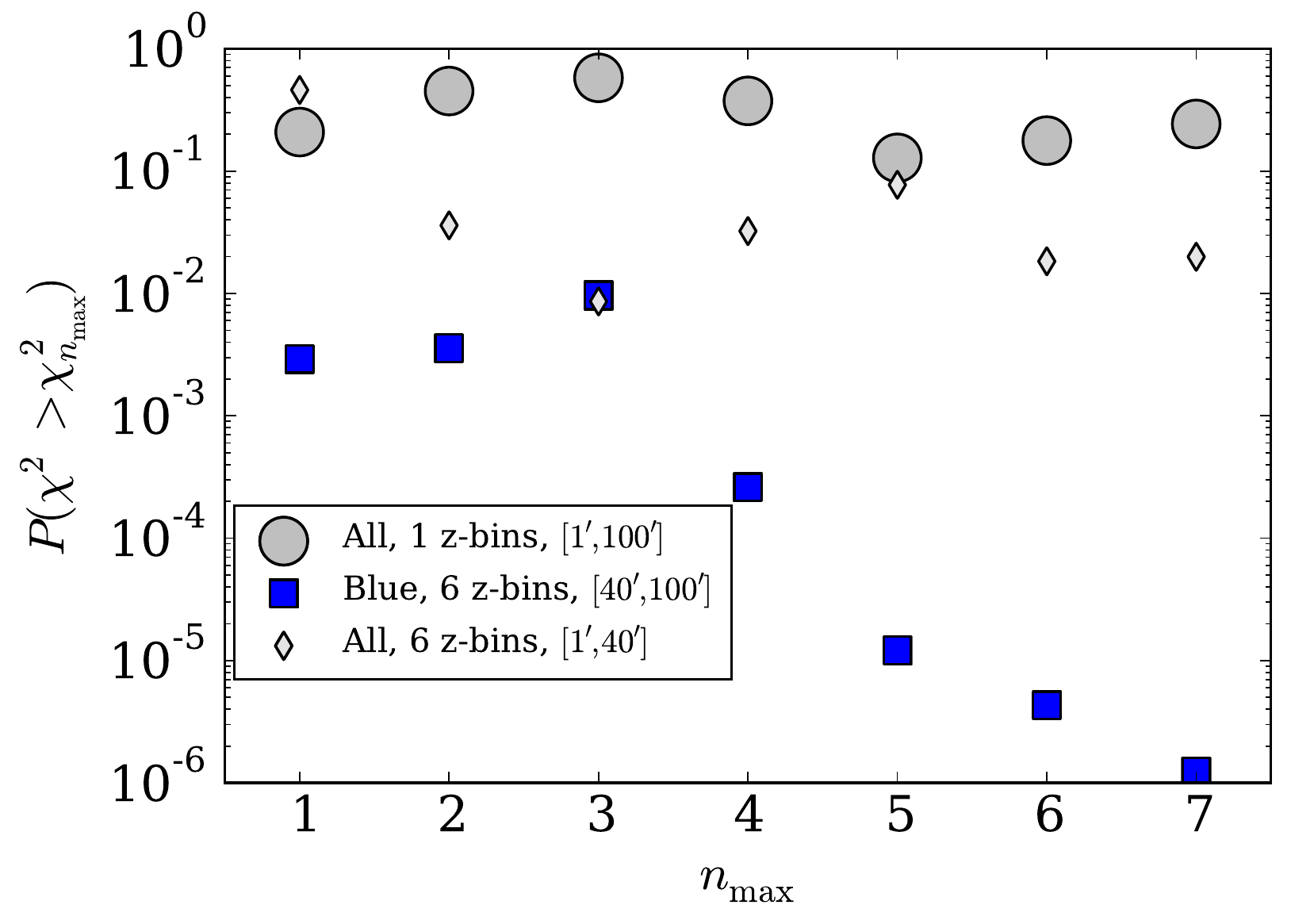}}
    \end{tabular}
    \caption{\small{P-values for $\chi^2$ { of B-mode compared to zero} versus the number of COSEBIs modes.
    $n_{\rm max}$ denotes the number of COSEBIs modes from $n=1$ to $n=n_{\rm max}$. 
    The p-value is the probability of the $\chi^2$ value being larger than the value found, 
    assuming $B_n=0$ is the model. A very small p-value shows a poor agreement between the theory and the estimated values.
    We reject the null hypothesis (zero B-modes) for p-values smaller than $0.01$, which corresponds to a significance larger than $99\%$. 
    The blue squares show the results for blue galaxies with 6 redshift bins for the largest angular scales,
    the light diamond belong to all galaxies with 6 redshift bins and small angular scales.
    Finally the grey circles show the p-values for all galaxies, a single redshift bin and the $[1',100']$ range.}}
    \label{fig:Pvalue}
  \end{center}
\end{figure}

\fig\ref{fig:Pvalue} shows the p-values for the COSEBIs $B_n$ versus $n_{\rm max}$, 
the maximum number of COSEBIs used starting from the first mode.   
The p-values are shown for the three configurations which are closest to the previous CFHTLenS analysis described in \sect\ref{sec:AnalysisSetups}.
The grey circles correspond to $[1',100']$ range without tomography and with all galaxies, which resembles \cite{Kilbinger13}.
The blue squares belong to $[40', 100']$ angular range with tomography and blue galaxies similar to \cite{Kitching14}.
The diamonds configuration is the same as \cite{Heymans13}, 
where all galaxies in the angular range $[1',40']$ are considered and binned in redshift.
In this plot we see that the p-values for the single redshift bin case are always above $0.01$ which 
means that they are insignificant. 
In contrast, on large scales the B-modes are always below $0.01$ and are significant.
{ In addition, the tomographic analysis using the lower angular range, $[1',40']$, also shows insignificant Bmodes with a p-value above $0.01$. }
When we use CCOSEBIs the B-modes significance decreases, as we will see in \tab\ref{tab:PvalueAll}.

\tab\ref{tab:PvalueAll} shows the p-values for all the cases that we have considered.
The first four columns indicate the set-up, while the last six show the p-values for that set-up for, 
$B_n=0$, $E_n=E_n^{\rm CFHTLenS}$, $E_n=E_n^{\rm Planck}$, 
$B^c=0$, $E^c=E^{\rm c, CFHTLenS}$ and $E^c=E^{\rm c, Planck}$, respectively. 
The p-values for CCOSEBIs are only shown for the tomographic cases where CCOSEBIs offers a compression. 
The $n_{\rm max}$ column shows the number of COSEBIs modes in each redshift bin which are used in the analysis. 
We show the results for both the first 2 and 7  COSEBIs. The p-values are written in boldface where they are larger than 
$0.01$ which corresponds to the significance level within $99\%$. 
Looking at the $B_n$ column and the single redshift bin cases, we see that the B-modes are only significant at large scales ($[40',100']$). 
{  When redshift binning is considered, with the exception of the  $[1',40']$ case there are significant B-modes in the data. }
The B-modes are not always consistent between the two galaxy populations 
which hints at a correlation between galaxy colour and residual systematics. 
Also notice that the largest scales show significant B-modes for all the different sets of data analysed.

The $B^c$ column shows the CCOSEBIs B-modes which are typically 
less significant than that of COSEBIs. As discussed before, this is due to the fact that the CCOSEBIs are based on linear 
combinations of COSEBIs which are most sensitive to cosmological parameters.  They are therefore not necessarily sensitive to the 
B-modes which, for CFHTLenS, appear to cancel to some degree with the compressed form of the statistic.
Consequently, to measure B-modes we need to use COSEBIs, which provide a complete set of functions for this analysis.

Comparing the $E_n^{\rm CFHTLenS}$ and $E_n^{\rm Planck}$ columns we see that Planck provides a better { match} to the 
single redshift bin data for all the cases\footnote{  Here we use p-values as a proxy for $\chi^2$ values, 
which would be used in sampling the parameter likelihood in a typical cosmological analysis. 
We will not attempt to reject either $E_n^{\rm CFHTLenS}$ or $E_n^{\rm Planck}$ using this method
or quantify their tension.}.
However, when tomography is considered the CFHTLenS cosmology provides a better { match}
with the exception of the very large scales. For blue galaxies at $[40', 100']$ the p-values 
for $E_n^{\rm CFHTLenS}$ and $E_n^{\rm Planck}$ are comparable. 
We also note that for many of the tomographic cases, neither provide a good { match}. 
When all galaxies are considered we need to add intrinsic alignment corrections to our model as was done in \cite{Heymans13}.
However, for blue galaxies the contribution from  intrinsic alignment is expected to be small, hence we expect and find a good fit to the
CFHTLenS values.

\begin{table*}
\caption{\small{P-values for $\chi^2$ { of} $B_n=0$, $E_n=E_n^{\rm CFHTLenS}$, $E_n=E_n^{\rm Planck}$, 
$B^c=0$, $E^c=E^{\rm c, CFHTLenS}$ and $E^c=E^{\rm c, Planck}$. 
The p-values denote the probability of the $\chi^2$ values being larger than the values found, 
assuming the model is correct.
Each row corresponds to a different angular range ($\theta$ range), 
group of galaxies (Galaxies), number of redshift bins (z-bins) and number of COSEBIs modes ($n_{\rm max}$) 
considered in the analysis.
The CCOSEBIs p-values are only shown for the tomographic case where the number
of CCOSEBIs modes is smaller than that of COSEBIs.
The P-values which are larger than 0.01 are shown in boldface and lie within the $99\%$ confidence limit.
See \tab\ref{tab:DOF} for the $\chi^2$ values and degrees-of-freedom for each entry in this table.}}
\begin{center}
\begin{adjustbox}{max width=\textwidth}

\begin{tabular}{ c  c  c  c  c  c  c  c  c  c }
&&&&\multicolumn{3}{c}{COSEBIs}&\multicolumn{3}{c}{CCOSEBIs}
\\ \hline
$\theta$ range & Galaxies & z-bins & $n_{\rm max}$&$ B_n$ & $E_n^{\rm CFHTLenS}$ & $E_n^{\rm Planck}$ 
 & $B^{\rm c}$ & $E^{\rm c, CFHTLenS}$ & $E^{\rm c, Planck}$ 
 \\ \hline
\multirow{8}{*}{$[1'-100']$}& \multirow{4}{*}{All}& \multirow{2}{*}{1}&2 & $\boldsymbol{4.5e-01}$ & $\boldsymbol{1.6e-01}$ & $\boldsymbol{3.4e-01}$ &$-$ &$-$ &$-$
\\ \cline{4-10}
& &&7 & $\boldsymbol{2.4e-01}$ & $\boldsymbol{4.5e-02}$ & $\boldsymbol{2.1e-01}$ &$-$ &$-$ &$-$
\\ \cline{4-10}

 \cline{3-10}
& & \multirow{2}{*}{6}&2 &$ 2.0e-03$ &$ 3.3e-03$ &$ 4.6e-04$ & $\boldsymbol{3.5e-02}$ & $\boldsymbol{7.8e-02}$ &$ \boldsymbol{1.1e-02}$
\\ \cline{4-10}
& &&7 &$ 6.6e-03$ &$ 6.1e-04$ &$ 2.0e-04$ & $\boldsymbol{8.0e-01}$ &$ 4.8e-04$ &$ 2.5e-04$
\\ \cline{4-10}

 \cline{3-10}

 \cline{2-10}
& \multirow{4}{*}{Blue}& \multirow{2}{*}{1}&2 & $\boldsymbol{2.1e-01}$ & $\boldsymbol{6.8e-02}$ & $\boldsymbol{3.5e-01}$ &$-$ &$-$ &$-$
\\ \cline{4-10}
& &&7 & $\boldsymbol{2.2e-01}$ &$ 2.1e-03$ & $\boldsymbol{1.3e-02}$ &$-$ &$-$ &$-$
\\ \cline{4-10}

 \cline{3-10}
& & \multirow{2}{*}{6}&2 &$ 4.4e-04$ & $\boldsymbol{3.5e-02}$ &$ 5.7e-03$ & $\boldsymbol{6.0e-02}$ & $\boldsymbol{2.7e-01}$ & $\boldsymbol{6.7e-02}$
\\ \cline{4-10}
& &&7 &$ 3.9e-03$ &$ 9.7e-03$ &$ 4.9e-03$ & $\boldsymbol{4.7e-01}$ & $\boldsymbol{1.1e-01}$ & $\boldsymbol{3.5e-02}$
\\ \cline{4-10}

 \cline{3-10}

 \cline{2-10}

\cline{1-10}
\multirow{8}{*}{$[1'-40']$}& \multirow{4}{*}{All}& \multirow{2}{*}{1}&2 & $\boldsymbol{7.4e-01}$ & $\boldsymbol{3.2e-02}$ & $\boldsymbol{5.1e-01}$ &$-$ &$-$ &$-$
\\ \cline{4-10}
& &&7 & $\boldsymbol{7.4e-01}$ & $\boldsymbol{1.2e-01}$ & $\boldsymbol{5.0e-01}$ &$-$ &$-$ &$-$
\\ \cline{4-10}

 \cline{3-10}
& & \multirow{2}{*}{6}&2 & $\boldsymbol{3.6e-02}$ &$ 3.0e-03$ &$ 1.3e-03$ & $\boldsymbol{2.5e-01}$ & $\boldsymbol{3.2e-02}$ &$ \boldsymbol{1.1e-02}$
\\ \cline{4-10}
& &&7 &$ \boldsymbol{2.0e-02}$ &$ 6.8e-03$ &$ 1.9e-03$ & $\boldsymbol{6.2e-01}$ &$ 4.2e-03$ &$ 1.3e-03$
\\ \cline{4-10}

 \cline{3-10}

 \cline{2-10}
& \multirow{4}{*}{Blue}& \multirow{2}{*}{1}&2 & $\boldsymbol{6.5e-01}$ & $\boldsymbol{1.8e-02}$ & $\boldsymbol{8.8e-01}$ &$-$ &$-$ &$-$
\\ \cline{4-10}
& &&7 & $\boldsymbol{2.7e-01}$ &$ 3.2e-03$ & $\boldsymbol{2.4e-02}$ &$-$ &$-$ &$-$
\\ \cline{4-10}

 \cline{3-10}
& & \multirow{2}{*}{6}&2 & $\boldsymbol{3.8e-02}$ & $\boldsymbol{2.4e-02}$ & $\boldsymbol{1.2e-02}$ & $\boldsymbol{7.0e-01}$ & $\boldsymbol{3.2e-01}$ & $\boldsymbol{2.0e-01}$
\\ \cline{4-10}
& &&7 & $\boldsymbol{2.7e-01}$ &$ 8.9e-03$ &$ 2.6e-03$ & $\boldsymbol{7.6e-02}$ & $\boldsymbol{1.4e-01}$ & $\boldsymbol{5.7e-02}$
\\ \cline{4-10}

 \cline{3-10}

 \cline{2-10}

\cline{1-10}
\multirow{8}{*}{$[40'-100']$}& \multirow{4}{*}{All}& \multirow{2}{*}{1}&2 &$ 4.4e-03$ & $\boldsymbol{4.0e-02}$ & $\boldsymbol{6.9e-02}$ &$-$ &$-$ &$-$
\\ \cline{4-10}
& &&7 &$ 2.4e-03$ & $\boldsymbol{6.2e-02}$ & $\boldsymbol{8.7e-02}$ &$-$ &$-$ &$-$
\\ \cline{4-10}

 \cline{3-10}
& & \multirow{2}{*}{6}&2 &$ 1.1e-03$ &$ \boldsymbol{1.2e-02}$ & $\boldsymbol{1.6e-02}$ & $\boldsymbol{4.8e-02}$ &$ 4.5e-03$ &$ 5.9e-03$
\\ \cline{4-10}
& &&7 &$ 1.8e-06$ &$ 4.7e-06$ &$ 5.3e-06$ & $\boldsymbol{3.5e-02}$ &$ 6.4e-03$ &$ 8.9e-03$
\\ \cline{4-10}

 \cline{3-10}

 \cline{2-10}
& \multirow{4}{*}{Blue}& \multirow{2}{*}{1}&2 &$ 5.5e-03$ & $\boldsymbol{1.5e-01}$ & $\boldsymbol{2.1e-01}$ &$-$ &$-$ &$-$
\\ \cline{4-10}
& &&7 &$ 2.9e-03$ & $\boldsymbol{4.4e-02}$ & $\boldsymbol{5.5e-02}$ &$-$ &$-$ &$-$
\\ \cline{4-10}

 \cline{3-10}
& & \multirow{2}{*}{6}&2 &$ 3.6e-03$ & $\boldsymbol{6.7e-02}$ & $\boldsymbol{7.4e-02}$ &$ 7.3e-04$ & $\boldsymbol{1.1e-01}$ & $\boldsymbol{1.2e-01}$
\\ \cline{4-10}
& &&7 &$ 1.2e-06$ &$ 1.1e-04$ &$ 1.1e-04$ &$ 9.6e-04$ & $\boldsymbol{1.7e-01}$ & $\boldsymbol{1.8e-01}$
\\ \cline{4-10}

 \cline{3-10}

 \cline{2-10}

\cline{1-10}
\end{tabular}

\end{adjustbox}
\end{center}
\label{tab:PvalueAll}
\end{table*}

\begin{table*}

\end{table*}

\subsection{Single Parameter Fit}

We use a very simple parametrization to fit the theory to data, consisting of one free parameter. 
We find its best fit value by minimizing its $\chi^2$ value and the error to the fit corresponds to the parameter value at
$\Delta \chi^2=1$ around the minimum $\chi^2$. 
% { Here we use the \cite{Hartlap07} correction, for the covariances, which provides a good approximation,
% since we are not interested in the tails of the distribution and the maximum value is not affected by the inaccuracies in the 
% estimated covariance matrices.}
For the B-modes the single parameter model we use is a constant, 
\begin{equation}
 B_n=K_B\;, ~~~~~~~~~{\rm and} ~~~~~~~~~B^{\rm c}=K_{B^{\rm c}}\;,
\end{equation}
 for COSEBIs and CCOSEBIs, respectively. 
For the E-modes the models are a constant, times the theory E-modes, with CFHTLenS+WMAP7 and 
Planck cosmologies. For COSEBIs these are 
\begin{align}
 E_n&=K_{\rm E}^{\rm CFHTLenS}E_n^{\rm CFHTLenS}\;,
 \end{align}
 and 
 \begin{align}
E_n&=K_{\rm E}^{\rm Planck} E_n^{\rm Planck}\;,
\end{align}
whereas for CCOSEBIs 
\begin{equation}
 E^{\rm c}=K_{\rm E^{\rm c}}^{\rm CFHTLenS}E^{\rm c, CFHTLenS},
\end{equation}
and 
\begin{equation}
 E^{\rm c}=K_{\rm E^{\rm c}}^{\rm Planck}E^{\rm c, Planck}\;,
\end{equation}
are the two models. 
The best fit and error values for $K_{\rm B}$,  $K_{\rm E}^{\rm CFHTLenS}$, 
$K_{\rm E}^{\rm Planck}$, $K_{\rm B^{\rm c}}$,  $K_{\rm E^{\rm c}}^{\rm CFHTLenS}$, $K_{\rm E^{\rm c}}^{\rm Planck}$
are listed in \tab\ref{tab:Ks}. The format of this table is the same as \tab\ref{tab:PvalueAll}. 
Null B-modes result in a statistically zero $K_{\rm B}$,
however, a statistically zero $K_{\rm B}$ is not a sufficient condition for B-modes to be zero. 
The rows for which the COSEBIs B-modes are consistent with zero from the p-value test are shown in boldface. 
Some of the $K_{\rm B}$ values which are consistent with zero in this table correspond to significant B-modes 
from the p-value test. This shows that the B-mode pattern in the data is not always well-modelled by a constant value. 

\begin{table*}
\caption{\small{Best fit values for $K_{\rm B}$,  $K_{\rm E}^{\rm CFHTLenS}$, 
$K_{\rm E}^{\rm Planck}$, $K_{\rm B^{\rm c}}$, 
$K_{\rm E^{\rm c}}^{\rm CFHTLenS}$, $K_{\rm E^{\rm c}}^{\rm Planck}$.
We use a parameter, $K_{\rm X}$, to fit to the models given by 
$B_n=K_B$, $E_n=K_{\rm E}^{\rm CFHTLenS}E_n^{\rm CFHTLenS}$, 
$E_n=K_{\rm E}^{\rm Planck} E_n^{\rm Planck}$
,$B^{\rm c}=K_{B^{\rm c}}$, $E^{\rm c}=K_{\rm E^{\rm c}}^{\rm CFHTLenS}E^{\rm c, CFHTLenS}$ 
and $E^{\rm c}=K_{\rm E^{\rm c}}^{\rm Planck}E^{\rm c, Planck}$. 
The errors on the fitted values show the $\Delta \chi^2=1$ value for the fit parameters.
The CCOSEBIs values are only shown for the tomographic 
case where the number of CCOSEBIs modes is smaller than that of COSEBIs.
The cases for which the p-values for null B-modes>0.01 are shown in boldface. 
The first column shows the angular range considered, while the following
columns show the galaxies used, the number of redshift bins and the number of COSEBIs modes in the analysis.}}
\begin{center}
\begin{adjustbox}{max width=\textwidth}
\begin{tabular}{ c  c  c  c  c  c  c  c  c  c }
&&&&\multicolumn{3}{c}{COSEBIs}&\multicolumn{3}{c}{CCOSEBIs}
\\ \hline
$\theta$ range & Galaxies & z-bins & $n_{\rm max}$& $K_{\rm B}\times10^{11}$ & $K_{\rm E}^{\rm CFHTLenS}$ & $K_{\rm E}^{\rm Planck}$ 
 & $K_{\rm B^{\rm c}}\times10^{15}$ & $K_{\rm E^{\rm c}}^{\rm CFHTLenS}$ & $K_{\rm E^{\rm c}}^{\rm Planck}$ 
 \\ \hline
\multirow{8}{*}{$[1'-100']$}& \multirow{4}{*}{All}& \multirow{2}{*}{1}&2 & $\boldsymbol{1.70\pm 1.39}$ & $\boldsymbol{1.32\pm 0.20}$ & $\boldsymbol{0.84\pm 0.13}$ &$-$ &$-$ &$-$
\\ \cline{4-10}
& &&7 & $\boldsymbol{0.63\pm 1.29}$ & $\boldsymbol{1.41\pm 0.19}$ & $\boldsymbol{0.89\pm 0.12}$ &$-$ &$-$ &$-$
\\ \cline{4-10}

 \cline{3-10}
& & \multirow{2}{*}{6}&2 & $1.80\pm 1.38$ & $1.04\pm 0.17$ & $0.68\pm 0.11$ & $0.93\pm 13.36$ & $1.08\pm 0.17$ & $0.70\pm 0.11$
\\ \cline{4-10}
& &&7 & $0.53\pm 1.26$ & $1.06\pm 0.16$ & $0.69\pm 0.10$ & $26.13\pm 28.48$ & $1.22\pm 0.16$ & $0.79\pm 0.10$
\\ \cline{4-10}

 \cline{3-10}

 \cline{2-10}
& \multirow{4}{*}{Blue}& \multirow{2}{*}{1}&2 & $\boldsymbol{1.91\pm 1.60}$ & $\boldsymbol{1.38\pm 0.21}$ & $\boldsymbol{0.88\pm 0.13}$ &$-$ &$-$ &$-$
\\ \cline{4-10}
& &&7 & $\boldsymbol{0.91\pm 1.50}$ & $\boldsymbol{1.41\pm 0.19}$ & $\boldsymbol{0.90\pm 0.12}$ &$-$ &$-$ &$-$
\\ \cline{4-10}

 \cline{3-10}
& & \multirow{2}{*}{6}&2 & $1.23\pm 1.63$ & $0.99\pm 0.17$ & $0.65\pm 0.11$ & $-6.64\pm 5.74$ & $1.07\pm 0.17$ & $0.70\pm 0.11$
\\ \cline{4-10}
& &&7 & $0.65\pm 1.50$ & $1.09\pm 0.16$ & $0.71\pm 0.10$ & $22.98\pm 18.93$ & $1.15\pm 0.16$ & $0.74\pm 0.10$
\\ \cline{4-10}

 \cline{3-10}

 \cline{2-10}

\cline{1-10}
\multirow{8}{*}{$[1'-40']$}& \multirow{4}{*}{All}& \multirow{2}{*}{1}&2 & $\boldsymbol{0.03\pm 0.65}$ & $\boldsymbol{1.45\pm 0.19}$ & $\boldsymbol{0.91\pm 0.12}$ &$-$ &$-$ &$-$
\\ \cline{4-10}
& &&7 & $\boldsymbol{-0.48\pm 0.57}$ & $\boldsymbol{1.43\pm 0.19}$ & $\boldsymbol{0.90\pm 0.12}$ &$-$ &$-$ &$-$
\\ \cline{4-10}

 \cline{3-10}
& & \multirow{2}{*}{6}&2 & $\boldsymbol{-0.48\pm 0.65}$ & $\boldsymbol{1.15\pm 0.16}$ & $\boldsymbol{0.74\pm 0.10}$ & $\boldsymbol{-3.14\pm 2.87}$ & $\boldsymbol{1.16\pm 0.16}$ & $\boldsymbol{0.75\pm 0.10}$
\\ \cline{4-10}
& &&7 & $\boldsymbol{-0.63\pm 0.55}$ & $\boldsymbol{0.99\pm 0.16}$ & $\boldsymbol{0.64\pm 0.10}$ & $\boldsymbol{-2.44\pm 12.31}$ & $\boldsymbol{1.15\pm 0.16}$ & $\boldsymbol{0.75\pm 0.10}$
\\ \cline{4-10}

 \cline{3-10}

 \cline{2-10}
& \multirow{4}{*}{Blue}& \multirow{2}{*}{1}&2 & $\boldsymbol{-0.57\pm 0.78}$ & $\boldsymbol{1.53\pm 0.20}$ & $\boldsymbol{0.96\pm 0.13}$ &$-$ &$-$ &$-$
\\ \cline{4-10}
& &&7 & $\boldsymbol{-1.34\pm 0.68}$ & $\boldsymbol{1.45\pm 0.20}$ & $\boldsymbol{0.92\pm 0.12}$ &$-$ &$-$ &$-$
\\ \cline{4-10}

 \cline{3-10}
& & \multirow{2}{*}{6}&2 & $\boldsymbol{-0.94\pm 0.77}$ & $\boldsymbol{1.18\pm 0.17}$ & $\boldsymbol{0.76\pm 0.11}$ & $\boldsymbol{0.46\pm 2.83}$ & $\boldsymbol{1.19\pm 0.17}$ & $\boldsymbol{0.77\pm 0.11}$
\\ \cline{4-10}
& &&7 & $\boldsymbol{-1.61\pm 0.67}$ & $\boldsymbol{1.00\pm 0.16}$ & $\boldsymbol{0.65\pm 0.10}$ & $\boldsymbol{15.47\pm 8.86}$ & $\boldsymbol{1.16\pm 0.17}$ & $\boldsymbol{0.75\pm 0.11}$
\\ \cline{4-10}

 \cline{3-10}

 \cline{2-10}

\cline{1-10}
\multirow{8}{*}{$[40'-100']$}& \multirow{4}{*}{All}& \multirow{2}{*}{1}&2 & $-8.76\pm 3.36$ & $3.12\pm 1.29$ & $2.17\pm 0.90$ &$-$ &$-$ &$-$
\\ \cline{4-10}
& &&7 & $-2.25\pm 2.05$ & $3.04\pm 1.27$ & $2.12\pm 0.89$ &$-$ &$-$ &$-$
\\ \cline{4-10}

 \cline{3-10}
& & \multirow{2}{*}{6}&2 & $-7.61\pm 3.38$ & $3.39\pm 1.11$ & $2.39\pm 0.79$ & $-281.32\pm 316.00$ & $2.74\pm 1.13$ & $1.92\pm 0.81$
\\ \cline{4-10}
& &&7 & $1.51\pm 1.99$ & $2.68\pm 1.09$ & $1.84\pm 0.77$ & $-320.73\pm 557.68$ & $3.17\pm 1.17$ & $2.24\pm 0.83$
\\ \cline{4-10}

 \cline{3-10}

 \cline{2-10}
& \multirow{4}{*}{Blue}& \multirow{2}{*}{1}&2 & $-9.79\pm 4.04$ & $2.77\pm 1.45$ & $1.94\pm 1.02$ &$-$ &$-$ &$-$
\\ \cline{4-10}
& &&7 & $-1.99\pm 2.44$ & $2.57\pm 1.44$ & $1.80\pm 1.01$ &$-$ &$-$ &$-$
\\ \cline{4-10}

 \cline{3-10}
& & \multirow{2}{*}{6}&2 & $-8.05\pm 3.97$ & $2.12\pm 1.20$ & $1.52\pm 0.85$ & $-222.44\pm 208.12$ & $2.04\pm 1.20$ & $1.47\pm 0.86$
\\ \cline{4-10}
& &&7 & $1.91\pm 2.37$ & $1.40\pm 1.21$ & $0.98\pm 0.86$ & $412.94\pm 508.33$ & $1.81\pm 1.23$ & $1.31\pm 0.88$
\\ \cline{4-10}

 \cline{3-10}

 \cline{2-10}

\cline{1-10}
\end{tabular}

\end{adjustbox}
\end{center}
\label{tab:Ks}
\end{table*}

\subsection{Comparison to previous analyses}

The Tables \ref{tab:PvalueAll} and \ref{tab:Ks} allow us to compare 
our results with the previous CFHTLenS cosmic shear analysis.
We first consider \cite{Heymans12} who detail a systematics test using an E-B mode decomposition 
for three different two-point statistics; the top-hat shear variance, the 2PCF and the mass aperture statistics.  
Analysing angular scales from $[1',60']$ applying no redshift binning, they found no significant B-modes, which 
is consistent with our results.

\cite{Kilbinger13} performed a two-dimensional analysis of the data using several cosmic shear
methods, including the COSEBIs.
The aim of their work was to use a large angular range to estimate cosmological parameters,
but they faced difficulties estimating the COSEBIs from their mock data, 
known as the Clone simulations.
The main reason for their difficulties was the fact that the accuracy 
of the simulations for very large angular scales is limited, due to the finite box size. 
Consequently, they did not use COSEBIs for their final analysis of the data. 
Here we used updated simulations (SLICS) with better accuracy for large scales and did not encounter similar problems. 
We compare their results with our $[1',100']$ angular range with all galaxies and a single redshift bin. 
They reported insignificant B-modes which is consistent with our results.

\cite{Kitching14} restricted their study to large scales and blue galaxies with redshift information.
They reported no significant B-modes. 
Although the scales they used are defined in Fourier space where they performed their analysis, they roughly correspond to 
the large scales that we have considered here. 
In contrast to their study we find very significant B-modes in the $[40', 100']$ range.
One reason for this inconsistency could be that their mask model lacks the precision to 
find the B-modes \citep[see][for mask modelling]{Asgari16}. 
In contrast to power spectrum analysis, mask modelling has little or no effect on the estimation of COSEBIs.
{ Alternatively, this inconsistency could be due to the complexity of translating
the angular ranges used in a COSEBIs analysis to the Fourier modes considered in 3D-lensing.}

Our best fit CFHTLenS+WMAP7 fiducial cosmology comes from \cite{Heymans13}, who used the $[1',40']$ range with 
tomography.  They did not incorporate any E/B-mode decomposition methods in their analysis
 since they used 2PCFs to find their best fitting values. 
For the angular range they used, we find significant B-modes when all galaxies and 7 COSEBIs modes are considered. 
When only 2 COSEBIs modes are considered, or only using blue galaxies, the B-modes are consistent with zero. 
Considering blue galaxies only where intrinsic alignments are not important, 
we see that our measurements favour their best fit values in comparison to Planck.
In particular, the CCOSEBIs { matches} 
to both theoretical values for this case, however the CFHTLenS+WMAP7 { is a better match},
as expected. Aside from our choice of observables 
{ and the modelling of the intrinsic alignments, there are no} other differences between our study and \cite{Heymans13}.

\cite{Fu14} added three-point statistics to the \cite{Kilbinger13} 2D analysis and found significant B-modes in their 
third order statistics. Our findings combined with theirs show that there are still (high-order) residual systematic errors left in the CFHTLenS
data.

\section{Conclusions}
\label{sec:conclusions}
% summary, E-B mode correlation. power spectrum pipeline differences. PSF ellipticity and shear correlations. 

In this paper we revisited the CFHTLenS data 
and found evidence for systematic errors on large scales, and when
the data is analysed in tomographic bins. 
We used COSEBIs, which is a robust efficient and complete method for E/B-mode separation.  
We expect weak lensing to predominantly produce E-modes, making B-modes undesirable. 
% Most of the B-mode contribution typically comes from remaining systematics from the data reduction procedure. 
Although the absence of B-modes does not guarantee a perfect data analysis, 
it is a necessary condition for a survey like CFHTLenS.
For future large scale and space based surveys, where the measurement errors are significantly smaller,
the B-modes could also indicate other physical phenomena. For example we know that some intrinsic  alignment
models predict these modes \citep[see][and references therein]{Blazek11}. 
Before performing our analysis we carried out a number of blind tests on cosmological simulations, to test the 
accuracy of our pipelines which are reported in the Appendix. 
The significance of the B-modes we found is highest for large scales, $[40', 100']$,
especially when the galaxies are divided into redshift bins. 
They also depend on the galaxy population used in the analysis. 
We repeated our analysis for blue and all galaxies, 
since blue galaxies do not show a strong intrinsic alignment signal.

Our COSEBIs measurement on tomographic data is the first of its kind. 
Previously, all COSEBIs data analysis has been limited to 2D cosmic shear data.
Dividing galaxies into different redshift bins tightens the constraints on cosmological parameters, as it adds 
information about structure evolution, which is essential for constraining dark energy parameters.
Adding redshift information to data analysis increases the total number of COSEBIs that need to 
be measured.  This makes covariance matrix estimations more challenging, since  a larger number of simulations are needed 
to reach a satisfactory precision. To alleviate this we showed the first measurement of compressed COSEBIs (CCOSEBIs) 
which are composed of linear combinations of the COSEBIs 
that are most sensitive to cosmological information. 
This compression reduces the number of observables substantially. 
In this study where we used 7 COSEBIs modes and 6 redshift bins.
The total number of 147 COSEBIs modes, reduced to 20 CCOSEBIs to estimate 5 cosmological parameters. 
As a result, the estimated covariance for the CCOSEBIs has a higher precision. 

We analysed our data according to angular scale and galaxy type, 
as well as analysing the data with and without tomographic redshift bins.
We considered different samples of galaxies to compare our results 
with previous cosmic shear analyses of the CFHTLenS data.
Since our analysis focuses on tests for systematic errors, 
instead of parameter estimation, we compare our measurements to two different
flat $\Lambda$CDM cosmological models with Planck and CFHTLenS+WMAP7 best fit parameters. 
We calculated the goodness-of-fit of these two models to our data for the full range of analyses. 
The figure-of-merit used the p-value for a $\chi^2$ analysis, 
which indicates the probability of finding a $\chi^2$ value larger than the value found. 
We highlighted the values which corresponded to at least $99\%$ confidence. 
In addition, we used the same method to report the significance of the B-modes we found. 
We also used a simple parametrization to find the best fit values of a single parameter to our data. 
{  We find consistent results with \cite{Heymans13}, with little or insignificant B-modes in $[1',40']$ range. 
Both COSEBIs and CCOSEBIs show a better match to CFHTLenS cosmology over Planck for 
this angular range with redshift binning. }

We compared our large scale results with blue galaxies and tomography with \cite{Kitching14}, 
were we found the most significant B-mode signal, which is in tension with their finding of a zero B-mode. 
Our measured E-modes for this configuration 
{ show a slightly better match to Planck cosmology} in agreement with \cite{Kitching14}. 
However, the cosmic shear information in this range is the lowest and as a result it has the weakest constraining power.
On the other hand, as the modelling of baryons is associated with a rather
large uncertainty, restricting the analysis to larger angular scales
most likely removes systematics due to modelling.

Our results are in line with \cite{Kilbinger13} 
who reported insignificant B-modes for a single redshift bin and a wide angular range with all galaxies.
Our COSEBIs measurements { in this range match with Planck cosmology better than CFHTLenS+WMAP7} which 
is also consistent with the parameter constraints from \cite{Kilbinger13} 
which are in less tension with Planck than the tomographic CFHTLenS analysis. 
%We note that there are differences between 
%the power spectrum pipelines used here and in the other analysis and our pipeline does not account 
%for effects of baryons beyond the shape parameter.

\cite{Fu14} have also reported B-modes in CFHTLenS for three-point statistics, using aperture mass statistics. 
Our results together show that there are remaining systematics left in the data.  One hypothesis is that these  
systematics arise from selection effects which introduce
a correlation between the PSF ellipticity and the galaxy ellipticity when galaxies are divided into redshift bins.
The B-modes we measured are in general larger when tomography is considered.
This will be investigated in more detail in our future work.
Although not quantified here we can see an anti-correlation between the E-modes and B-modes which is visible in the plots. 
This suggests that the systematic errors affect the E/B-modes in the same way. 

The CCOSEBIs show insignificant B-modes for many cases that we have studied, 
even when the COSEBIs indicate otherwise. Currently, we do not have a full understanding of how systematic errors affect
E/B-modes. This can be investigated by simulating systematic errors that show a similar signature to the ones found here,
and examine their effect on E-modes. 
Nevertheless, assuming that the systematic errors affect the two modes in the same way (as hinted by the E/B anti-correlation),
 we can conclude that for the cases where the CCOSEBIs B-modes are negligible, they are
not degenerate with the cosmological parameters, which means that the B-modes 
detected by the COSEBIs should not bias the parameter estimations. 
Note that given the assumption above, if the CCOSEBIs modes are not sensitive 
to the systematics then they must be orthogonal to them.

% The same cannot be said for the $[40', 100']$ case where the CCOSEBIs show a significant B-modes as well. 

% In order to understand the appearance of B-modes when the data was separated into tomographic bins, we analysed the 
% correlation between PSF ellipticity and galaxy ellipticity as a function of the measured best-fit photometric redshift.
% We found that different photometric redshift slices had a preferred direction of the galaxy with respect to the PSF, which
% on average across all redshifts was consistent with zero.  
% This could perhaps indicate that tomographic binning introduces an intrinsic ellipticity
% selection bias which may be the root of the B-modes observed in the data. 
% This will be explored further in future work.

The methods and pipelines used in this analysis can and should be used with any other reduced cosmic shear dataset. 
COSEBIs is arguably the best method for testing for B-modes in the data
and can also be used for measuring B-modes for cosmological analysis. Future and ongoing surveys will suffer more 
from inaccuracies in their covariance estimations, which can be remedied by either using a larger number of simulations or 
by decreasing the number of data points used in the analysis. As N-body simulations are expensive and time consuming we 
recommend using the compression method in \cite{AS2014} applied to COSEBIs to find CCOSEBIs
and accurate cosmological parameters.

\section*{Acknowledgements}
MA and CH acknowledge support from the European Research Council through grant number 240185 and 647112.
CB acknowledges the support of the Australian Research Council through the award of a Future Fellowship.
PS is supported by the Deutsche Forschungsgemeinschaft under the program TR33 and SCHN-342/13.
JHD acknowledges support from NSERC. 
Computations for the $N$-body simulations were performed on the GPC supercomputer at the SciNet HPC Consortium. 
SciNet is funded by: the Canada Foundation for Innovation under the auspices of Compute Canada; the Government of Ontario; 
Ontario Research Fund - Research Excellence; and the University of Toronto.
We thank the CFHTLenS team for making the data publicly available.
This work is based on observations obtained with MegaPrime/MegaCam, a joint project of CFHT and CEA/IRFU, 
at the Canada-France-Hawaii Telescope (CFHT) which is operated by the National Research Council (NRC) of Canada, 
the Institut National des Sciences de l'Univers of 
the Centre National de la Recherche Scientifique (CNRS) of France, and the University of Hawaii. 
This research used the facilities of the Canadian Astronomy Data Centre operated by 
the National Research Council of Canada with the support of the Canadian Space Agency. 
CFHTLenS data processing was made possible thanks to significant computing support 
from the NSERC Research Tools and Instruments grant program.
%%%%%%%%%%%%%%%%%%%% REFERENCES %%%%%%%%%%%%%%%%%%

% The best way to enter references is to use BibTeX:

 \bibliographystyle{mnras}
 \bibliography{COSEBIs}

\begin{thebibliography}{}
\makeatletter
\relax
\def\mn@urlcharsother{\let\do\@makeother \do\$\do\&\do\#\do\^\do\_\do\%\do\~}
\def\mn@doi{\begingroup\mn@urlcharsother \@ifnextchar [ {\mn@doi@}
  {\mn@doi@[]}}
\def\mn@doi@[#1]#2{\def\@tempa{#1}\ifx\@tempa\@empty \href
  {http://dx.doi.org/#2} {doi:#2}\else \href {http://dx.doi.org/#2} {#1}\fi
  \endgroup}
\def\mn@eprint#1#2{\mn@eprint@#1:#2::\@nil}
\def\mn@eprint@arXiv#1{\href {http://arxiv.org/abs/#1} {{\tt arXiv:#1}}}
\def\mn@eprint@dblp#1{\href {http://dblp.uni-trier.de/rec/bibtex/#1.xml}
  {dblp:#1}}
\def\mn@eprint@#1:#2:#3:#4\@nil{\def\@tempa {#1}\def\@tempb {#2}\def\@tempc
  {#3}\ifx \@tempc \@empty \let \@tempc \@tempb \let \@tempb \@tempa \fi \ifx
  \@tempb \@empty \def\@tempb {arXiv}\fi \@ifundefined
  {mn@eprint@\@tempb}{\@tempb:\@tempc}{\expandafter \expandafter \csname
  mn@eprint@\@tempb\endcsname \expandafter{\@tempc}}}

\bibitem[\protect\citeauthoryear{{Addison}, {Huang}, {Watts}, {Bennett},
  {Halpern}, {Hinshaw}  \& {Weiland}}{{Addison} et~al.}{2016}]{Addison15}
{Addison} G.~E.,  {Huang} Y.,  {Watts} D.~J.,  {Bennett} C.~L.,  {Halpern} M.,
  {Hinshaw} G.,   {Weiland} J.~L.,  2016, \mn@doi [\apj]
  {10.3847/0004-637X/818/2/132}, \href
  {http://adsabs.harvard.edu/abs/2016ApJ...818..132A} {818, 132}

\bibitem[\protect\citeauthoryear{{Andrae}, {Schulze-Hartung}  \&
  {Melchior}}{{Andrae} et~al.}{2010}]{Andrea10}
{Andrae} R.,  {Schulze-Hartung} T.,   {Melchior} P.,  2010, preprint, \href
  {http://adsabs.harvard.edu/abs/2010arXiv1012.3754A} {} (\mn@eprint {arXiv}
  {1012.3754})

\bibitem[\protect\citeauthoryear{{Asgari} \& {Schneider}}{{Asgari} \&
  {Schneider}}{2015}]{AS2014}
{Asgari} M.,  {Schneider} P.,  2015, \mn@doi [\aap]
  {10.1051/0004-6361/201424905}, \href
  {http://adsabs.harvard.edu/abs/2015A%26A...578A..50A} {578, A50}

\bibitem[\protect\citeauthoryear{{Asgari}, {Schneider}  \& {Simon}}{{Asgari}
  et~al.}{2012}]{Asgari12}
{Asgari} M.,  {Schneider} P.,   {Simon} P.,  2012, \mn@doi [A\&A]
  {10.1051/0004-6361/201218828}, \href
  {http://adsabs.harvard.edu/abs/2012A%26A...542A.122A} {542, A122}

\bibitem[\protect\citeauthoryear{{Asgari}, {Taylor}, {Joachimi}  \&
  {Kitching}}{{Asgari} et~al.}{2016}]{Asgari16}
{Asgari} M.,  {Taylor} A.,  {Joachimi} B.,   {Kitching} T.~D.,  2016, preprint,
  \href {http://adsabs.harvard.edu/abs/2016arXiv161204664A} {} (\mn@eprint
  {arXiv} {1612.04664})

\bibitem[\protect\citeauthoryear{{Bartelmann} \& {Schneider}}{{Bartelmann} \&
  {Schneider}}{2001}]{BartelmannSchneider01}
{Bartelmann} M.,  {Schneider} P.,  2001, \mn@doi [\physrep]
  {10.1016/S0370-1573(00)00082-X}, \href
  {http://adsabs.harvard.edu/abs/2001PhR...340..291B} {340, 291}

\bibitem[\protect\citeauthoryear{{Battye} \& {Moss}}{{Battye} \&
  {Moss}}{2014}]{BattyeMoss14}
{Battye} R.~A.,  {Moss} A.,  2014, \mn@doi [Physical Review Letters]
  {10.1103/PhysRevLett.112.051303}, \href
  {http://adsabs.harvard.edu/abs/2014PhRvL.112e1303B} {112, 051303}

\bibitem[\protect\citeauthoryear{{Becker} et~al.,}{{Becker}
  et~al.}{2015}]{Becker15}
{Becker} M.~R.,  et~al., 2015, preprint, \href
  {http://adsabs.harvard.edu/abs/2015arXiv150705598B} {} (\mn@eprint {arXiv}
  {1507.05598})

\bibitem[\protect\citeauthoryear{{Benjamin} et~al.,}{{Benjamin}
  et~al.}{2013}]{Benjamin13}
{Benjamin} J.,  et~al., 2013, \mn@doi [\mnras] {10.1093/mnras/stt276}, \href
  {http://adsabs.harvard.edu/abs/2013MNRAS.431.1547B} {431, 1547}

\bibitem[\protect\citeauthoryear{{Blazek}, {McQuinn}  \& {Seljak}}{{Blazek}
  et~al.}{2011}]{Blazek11}
{Blazek} J.,  {McQuinn} M.,   {Seljak} U.,  2011, \mn@doi [\jcap]
  {10.1088/1475-7516/2011/05/010}, \href
  {http://adsabs.harvard.edu/abs/2011JCAP...05..010B} {5, 10}

\bibitem[\protect\citeauthoryear{{Bond} \& {Efstathiou}}{{Bond} \&
  {Efstathiou}}{1984}]{BondEfstathiou84}
{Bond} J.~R.,  {Efstathiou} G.,  1984, \mn@doi [\apjl] {10.1086/184362}, \href
  {http://adsabs.harvard.edu/abs/1984ApJ...285L..45B} {285, L45}

\bibitem[\protect\citeauthoryear{{Choi} et~al.,}{{Choi} et~al.}{2015}]{Choi15}
{Choi} A.,  et~al., 2015, preprint, \href
  {http://adsabs.harvard.edu/abs/2015arXiv151203626C} {} (\mn@eprint {arXiv}
  {1512.03626})

\bibitem[\protect\citeauthoryear{{Dossett}, {Ishak}, {Parkinson}  \&
  {Davis}}{{Dossett} et~al.}{2015}]{Dossett15}
{Dossett} J.~N.,  {Ishak} M.,  {Parkinson} D.,   {Davis} T.~M.,  2015, \mn@doi
  [\prd] {10.1103/PhysRevD.92.023003}, \href
  {http://adsabs.harvard.edu/abs/2015PhRvD..92b3003D} {92, 023003}

\bibitem[\protect\citeauthoryear{{Eifler}}{{Eifler}}{2011}]{Eifler11}
{Eifler} T.,  2011, \mn@doi [\mnras] {10.1111/j.1365-2966.2011.19502.x}, \href
  {http://adsabs.harvard.edu/abs/2011MNRAS.418..536E} {418, 536}

\bibitem[\protect\citeauthoryear{{Erben} et~al.,}{{Erben}
  et~al.}{2013}]{Erben13}
{Erben} T.,  et~al., 2013, \mn@doi [\mnras] {10.1093/mnras/stt928}, \href
  {http://adsabs.harvard.edu/abs/2013MNRAS.433.2545E} {433, 2545}

\bibitem[\protect\citeauthoryear{{Fu} \& {Kilbinger}}{{Fu} \&
  {Kilbinger}}{2010}]{Fu10}
{Fu} L.,  {Kilbinger} M.,  2010, \mn@doi [\mnras]
  {10.1111/j.1365-2966.2009.15720.x}, \href
  {http://adsabs.harvard.edu/abs/2010MNRAS.401.1264F} {401, 1264}

\bibitem[\protect\citeauthoryear{{Fu} et~al.,}{{Fu} et~al.}{2014}]{Fu14}
{Fu} L.,  et~al., 2014, \mn@doi [\mnras] {10.1093/mnras/stu754}, \href
  {http://adsabs.harvard.edu/abs/2014MNRAS.441.2725F} {441, 2725}

\bibitem[\protect\citeauthoryear{{Harnois-D{\'e}raps} \& {van
  Waerbeke}}{{Harnois-D{\'e}raps} \& {van Waerbeke}}{2015}]{Harnois15}
{Harnois-D{\'e}raps} J.,  {van Waerbeke} L.,  2015, \mn@doi [\mnras]
  {10.1093/mnras/stv794}, \href
  {http://adsabs.harvard.edu/abs/2015MNRAS.450.2857H} {450, 2857}

\bibitem[\protect\citeauthoryear{{Harnois-D{\'e}raps}, {van Waerbeke}, {Viola}
  \& {Heymans}}{{Harnois-D{\'e}raps} et~al.}{2015}]{HarnoisVanWaerbeke15}
{Harnois-D{\'e}raps} J.,  {van Waerbeke} L.,  {Viola} M.,   {Heymans} C.,
  2015, \mn@doi [\mnras] {10.1093/mnras/stv646}, \href
  {http://adsabs.harvard.edu/abs/2015MNRAS.450.1212H} {450, 1212}

\bibitem[\protect\citeauthoryear{{Hartlap}, {Simon}  \& {Schneider}}{{Hartlap}
  et~al.}{2007}]{Hartlap07}
{Hartlap} J.,  {Simon} P.,   {Schneider} P.,  2007, \mn@doi [\aap]
  {10.1051/0004-6361:20066170}, \href
  {http://adsabs.harvard.edu/abs/2007A%26A...464..399H} {464, 399}

\bibitem[\protect\citeauthoryear{{Heymans} et~al.,}{{Heymans}
  et~al.}{2012}]{Heymans12}
{Heymans} C.,  et~al., 2012, \mn@doi [\mnras]
  {10.1111/j.1365-2966.2012.21952.x}, \href
  {http://adsabs.harvard.edu/abs/2012MNRAS.427..146H} {427, 146}

\bibitem[\protect\citeauthoryear{{Heymans} et~al.,}{{Heymans}
  et~al.}{2013}]{Heymans13}
{Heymans} C.,  et~al., 2013, \mn@doi [\mnras] {10.1093/mnras/stt601}, \href
  {http://adsabs.harvard.edu/abs/2013MNRAS.432.2433H} {432, 2433}

\bibitem[\protect\citeauthoryear{{Hilbert}, {Hartlap}, {White}  \&
  {Schneider}}{{Hilbert} et~al.}{2009}]{Hilbert09}
{Hilbert} S.,  {Hartlap} J.,  {White} S.~D.~M.,   {Schneider} P.,  2009,
  \mn@doi [\aap] {10.1051/0004-6361/200811054}, \href
  {http://adsabs.harvard.edu/abs/2009A%26A...499...31H} {499, 31}

\bibitem[\protect\citeauthoryear{{Hildebrandt} et~al.,}{{Hildebrandt}
  et~al.}{2012}]{Hildebrandt12}
{Hildebrandt} H.,  et~al., 2012, \mn@doi [\mnras]
  {10.1111/j.1365-2966.2012.20468.x}, \href
  {http://adsabs.harvard.edu/abs/2012MNRAS.421.2355H} {421, 2355}

\bibitem[\protect\citeauthoryear{{Huff}, {Eifler}, {Hirata}, {Mandelbaum},
  {Schlegel}  \& {Seljak}}{{Huff} et~al.}{2014}]{Huff14}
{Huff} E.~M.,  {Eifler} T.,  {Hirata} C.~M.,  {Mandelbaum} R.,  {Schlegel} D.,
   {Seljak} U.,  2014, \mn@doi [\mnras] {10.1093/mnras/stu145}, \href
  {http://adsabs.harvard.edu/abs/2014MNRAS.440.1322H} {440, 1322}

\bibitem[\protect\citeauthoryear{{Joachimi} \& {Schneider}}{{Joachimi} \&
  {Schneider}}{2008}]{JoachimiSchneder08}
{Joachimi} B.,  {Schneider} P.,  2008, \mn@doi [\aap]
  {10.1051/0004-6361:200809971}, \href
  {http://adsabs.harvard.edu/abs/2008A%26A...488..829J} {488, 829}

\bibitem[\protect\citeauthoryear{{Joudaki} et~al.,}{{Joudaki}
  et~al.}{2016}]{Joudaki16}
{Joudaki} S.,  et~al., 2016, preprint (\mn@eprint {arXiv} {in prep})

\bibitem[\protect\citeauthoryear{{Kilbinger}}{{Kilbinger}}{2015}]{Kilbinger_review}
{Kilbinger} M.,  2015, \mn@doi [Reports on Progress in Physics]
  {10.1088/0034-4885/78/8/086901}, \href
  {http://adsabs.harvard.edu/abs/2015RPPh...78h6901K} {78, 086901}

\bibitem[\protect\citeauthoryear{{Kilbinger} et~al.,}{{Kilbinger}
  et~al.}{2013}]{Kilbinger13}
{Kilbinger} M.,  et~al., 2013, \mn@doi [\mnras] {10.1093/mnras/stt041}, \href
  {http://adsabs.harvard.edu/abs/2013MNRAS.430.2200K} {430, 2200}

\bibitem[\protect\citeauthoryear{{Kilbinger}, {Bonnett}  \&
  {Coupon}}{{Kilbinger} et~al.}{2014}]{KilbingerAthena14}
{Kilbinger} M.,  {Bonnett} C.,   {Coupon} J.,  2014, {athena: Tree code for
  second-order correlation functions}, Astrophysics Source Code Library
  (\mn@eprint {ascl} {1402.026})

\bibitem[\protect\citeauthoryear{{Kitching} et~al.,}{{Kitching}
  et~al.}{2014}]{Kitching14}
{Kitching} T.~D.,  et~al., 2014, \mn@doi [\mnras] {10.1093/mnras/stu934}, \href
  {http://adsabs.harvard.edu/abs/2014MNRAS.442.1326K} {442, 1326}

\bibitem[\protect\citeauthoryear{{K{\"o}hlinger}, {Viola}, {Valkenburg},
  {Joachimi}, {Hoekstra}  \& {Kuijken}}{{K{\"o}hlinger}
  et~al.}{2015}]{Kohlinger15}
{K{\"o}hlinger} F.,  {Viola} M.,  {Valkenburg} W.,  {Joachimi} B.,  {Hoekstra}
  H.,   {Kuijken} K.,  2015, preprint, \href
  {http://adsabs.harvard.edu/abs/2015arXiv150904071K} {} (\mn@eprint {arXiv}
  {1509.04071})

\bibitem[\protect\citeauthoryear{{Kuijken} et~al.,}{{Kuijken}
  et~al.}{2015}]{Kuijken15}
{Kuijken} K.,  et~al., 2015, \mn@doi [\mnras] {10.1093/mnras/stv2140}, \href
  {http://adsabs.harvard.edu/abs/2015MNRAS.454.3500K} {454, 3500}

\bibitem[\protect\citeauthoryear{{Laureijs} et~al.,}{{Laureijs}
  et~al.}{2011}]{EuclidRB11}
{Laureijs} R.,  et~al., 2011, ArXiv:1110.3193, \href
  {http://adsabs.harvard.edu/abs/2011arXiv1110.3193L} {}

\bibitem[\protect\citeauthoryear{{MacCrann}, {Zuntz}, {Bridle}, {Jain}  \&
  {Becker}}{{MacCrann} et~al.}{2015}]{MacCrann15}
{MacCrann} N.,  {Zuntz} J.,  {Bridle} S.,  {Jain} B.,   {Becker} M.~R.,  2015,
  \mn@doi [\mnras] {10.1093/mnras/stv1154}, \href
  {http://adsabs.harvard.edu/abs/2015MNRAS.451.2877M} {451, 2877}

\bibitem[\protect\citeauthoryear{{Mead}, {Peacock}, {Heymans}, {Joudaki}  \&
  {Heavens}}{{Mead} et~al.}{2015}]{Mead15}
{Mead} A.~J.,  {Peacock} J.~A.,  {Heymans} C.,  {Joudaki} S.,   {Heavens}
  A.~F.,  2015, \mn@doi [\mnras] {10.1093/mnras/stv2036}, \href
  {http://adsabs.harvard.edu/abs/2015MNRAS.454.1958M} {454, 1958}

\bibitem[\protect\citeauthoryear{{Melchior} \& {Viola}}{{Melchior} \&
  {Viola}}{2012}]{Melchior_Viola_2012}
{Melchior} P.,  {Viola} M.,  2012, \mn@doi [\mnras]
  {10.1111/j.1365-2966.2012.21381.x}, \href
  {http://adsabs.harvard.edu/abs/2012MNRAS.424.2757M} {424, 2757}

\bibitem[\protect\citeauthoryear{{Miller} et~al.,}{{Miller}
  et~al.}{2013}]{Miller13}
{Miller} L.,  et~al., 2013, \mn@doi [\mnras] {10.1093/mnras/sts454}, \href
  {http://adsabs.harvard.edu/abs/2013MNRAS.429.2858M} {429, 2858}

\bibitem[\protect\citeauthoryear{{Planck Collaboration} et~al.,}{{Planck
  Collaboration} et~al.}{2015a}]{Planck15}
{Planck Collaboration} et~al., 2015a, preprint, \href
  {http://adsabs.harvard.edu/abs/2015arXiv150201589P} {} (\mn@eprint {arXiv}
  {1502.01589})

\bibitem[\protect\citeauthoryear{{Planck Collaboration} et~al.,}{{Planck
  Collaboration} et~al.}{2015b}]{Planck15MG_DE}
{Planck Collaboration} et~al., 2015b, preprint, \href
  {http://adsabs.harvard.edu/abs/2015arXiv150201590P} {} (\mn@eprint {arXiv}
  {1502.01590})

\bibitem[\protect\citeauthoryear{{Press}, {Teukolsky}, {Vetterling}  \&
  {Flannery}}{{Press} et~al.}{2002}]{numerical}
{Press} W.~H.,  {Teukolsky} S.~A.,  {Vetterling} W.~T.,   {Flannery} B.~P.,
  2002, {Numerical Recipes in C++}.
Cambridge University Press

\bibitem[\protect\citeauthoryear{{Raveri}}{{Raveri}}{2015}]{Raveri15}
{Raveri} M.,  2015, preprint, \href
  {http://adsabs.harvard.edu/abs/2015arXiv151000688R} {} (\mn@eprint {arXiv}
  {1510.00688})

\bibitem[\protect\citeauthoryear{{Schneider} \& {Kilbinger}}{{Schneider} \&
  {Kilbinger}}{2007}]{SchneiderKilbinger07}
{Schneider} P.,  {Kilbinger} M.,  2007, \mn@doi [\aap]
  {10.1051/0004-6361:20065532}, \href
  {http://adsabs.harvard.edu/abs/2007A%26A...462..841S} {462, 841}

\bibitem[\protect\citeauthoryear{{Schneider}, {van Waerbeke}, {Jain}  \&
  {Kruse}}{{Schneider} et~al.}{1998}]{Shcneider98}
{Schneider} P.,  {van Waerbeke} L.,  {Jain} B.,   {Kruse} G.,  1998, \mn@doi
  [\mnras] {10.1046/j.1365-8711.1998.01422.x}, \href
  {http://adsabs.harvard.edu/abs/1998MNRAS.296..873S} {296, 873}

\bibitem[\protect\citeauthoryear{{Schneider}, {van Waerbeke}  \&
  {Mellier}}{{Schneider} et~al.}{2002}]{Schneider02}
{Schneider} P.,  {van Waerbeke} L.,   {Mellier} Y.,  2002, \mn@doi [\aap]
  {10.1051/0004-6361:20020626}, \href
  {http://adsabs.harvard.edu/abs/2002A%26A...389..729S} {389, 729}

\bibitem[\protect\citeauthoryear{{Schneider}, {Eifler}  \&
  {Krause}}{{Schneider} et~al.}{2010}]{SEK10}
{Schneider} P.,  {Eifler} T.,   {Krause} E.,  2010, \mn@doi [\aap]
  {10.1051/0004-6361/201014235}, \href
  {http://adsabs.harvard.edu/abs/2010A%26A...520A.116S} {520, A116}

\bibitem[\protect\citeauthoryear{{Sellentin} \& {Heavens}}{{Sellentin} \&
  {Heavens}}{2015}]{SellentinHeavens15}
{Sellentin} E.,  {Heavens} A.~F.,  2015, preprint, \href
  {http://adsabs.harvard.edu/abs/2015arXiv151105969S} {} (\mn@eprint {arXiv}
  {1511.05969})

\bibitem[\protect\citeauthoryear{{Semboloni}, {Hoekstra}, {Schaye}, {van
  Daalen}  \& {McCarthy}}{{Semboloni} et~al.}{2011}]{Semboloni11}
{Semboloni} E.,  {Hoekstra} H.,  {Schaye} J.,  {van Daalen} M.~P.,   {McCarthy}
  I.~G.,  2011, \mn@doi [\mnras] {10.1111/j.1365-2966.2011.19385.x}, \href
  {http://adsabs.harvard.edu/abs/2011MNRAS.417.2020S} {417, 2020}

\bibitem[\protect\citeauthoryear{{Semboloni}, {Hoekstra}  \&
  {Schaye}}{{Semboloni} et~al.}{2013}]{Semboloni13}
{Semboloni} E.,  {Hoekstra} H.,   {Schaye} J.,  2013, \mn@doi [\mnras]
  {10.1093/mnras/stt1013}, \href
  {http://adsabs.harvard.edu/abs/2013MNRAS.434..148S} {434, 148}

\bibitem[\protect\citeauthoryear{{Simpson} et~al.,}{{Simpson}
  et~al.}{2013}]{Simpson13}
{Simpson} F.,  et~al., 2013, \mn@doi [\mnras] {10.1093/mnras/sts493}, \href
  {http://adsabs.harvard.edu/abs/2013MNRAS.429.2249S} {429, 2249}

\bibitem[\protect\citeauthoryear{{Smith} et~al.,}{{Smith}
  et~al.}{2003}]{Smith03}
{Smith} R.~E.,  et~al., 2003, \mn@doi [\mnras]
  {10.1046/j.1365-8711.2003.06503.x}, \href
  {http://adsabs.harvard.edu/abs/2003MNRAS.341.1311S} {341, 1311}

\bibitem[\protect\citeauthoryear{{Spergel}, {Flauger}  \& {Hlo{\v
  z}ek}}{{Spergel} et~al.}{2015}]{Spergel15}
{Spergel} D.~N.,  {Flauger} R.,   {Hlo{\v z}ek} R.,  2015, \mn@doi [\prd]
  {10.1103/PhysRevD.91.023518}, \href
  {http://adsabs.harvard.edu/abs/2015PhRvD..91b3518S} {91, 023518}

\bibitem[\protect\citeauthoryear{{Takahashi}, {Sato}, {Nishimichi}, {Taruya}
  \& {Oguri}}{{Takahashi} et~al.}{2012}]{Takahashi12}
{Takahashi} R.,  {Sato} M.,  {Nishimichi} T.,  {Taruya} A.,   {Oguri} M.,
  2012, \mn@doi [\apj] {10.1088/0004-637X/761/2/152}, \href
  {http://adsabs.harvard.edu/abs/2012ApJ...761..152T} {761, 152}

\bibitem[\protect\citeauthoryear{{Taylor} \& {Joachimi}}{{Taylor} \&
  {Joachimi}}{2014}]{TaylorJoachimi14}
{Taylor} A.,  {Joachimi} B.,  2014, \mn@doi [\mnras] {10.1093/mnras/stu996},
  \href {http://adsabs.harvard.edu/abs/2014MNRAS.442.2728T} {442, 2728}

\bibitem[\protect\citeauthoryear{{The Dark Energy Survey Collaboration}
  et~al.,}{{The Dark Energy Survey Collaboration} et~al.}{2015}]{DES15_CP}
{The Dark Energy Survey Collaboration} et~al., 2015, preprint, \href
  {http://adsabs.harvard.edu/abs/2015arXiv150705552T} {} (\mn@eprint {arXiv}
  {1507.05552})

\bibitem[\protect\citeauthoryear{{Velander} et~al.,}{{Velander}
  et~al.}{2014}]{Velander14}
{Velander} M.,  et~al., 2014, \mn@doi [\mnras] {10.1093/mnras/stt2013}, \href
  {http://adsabs.harvard.edu/abs/2014MNRAS.437.2111V} {437, 2111}

\bibitem[\protect\citeauthoryear{{Verde}, {Feeney}, {Mortlock}  \&
  {Peiris}}{{Verde} et~al.}{2013}]{Verde13}
{Verde} L.,  {Feeney} S.~M.,  {Mortlock} D.~J.,   {Peiris} H.~V.,  2013,
  \mn@doi [\jcap] {10.1088/1475-7516/2013/09/013}, \href
  {http://adsabs.harvard.edu/abs/2013JCAP...09..013V} {9, 13}

\makeatother
\end{thebibliography}

%%%%%%%%%%%%%%%%% APPENDICES %%%%%%%%%%%%%%%%%%%%%

\appendix

\section{Pipeline Optimization and Verification}

In this Appendix we present a series of tests to explore 
the effects of noise and discrete integration on the determination of the COSEBIs for CFHTLenS-like data. 

\subsection{Power spectra vs. 2PCFs}
\label{app:PvsKsi}

\begin{figure}
  \begin{center}
    \begin{tabular}{c}
      \resizebox{80mm}{!}{\includegraphics{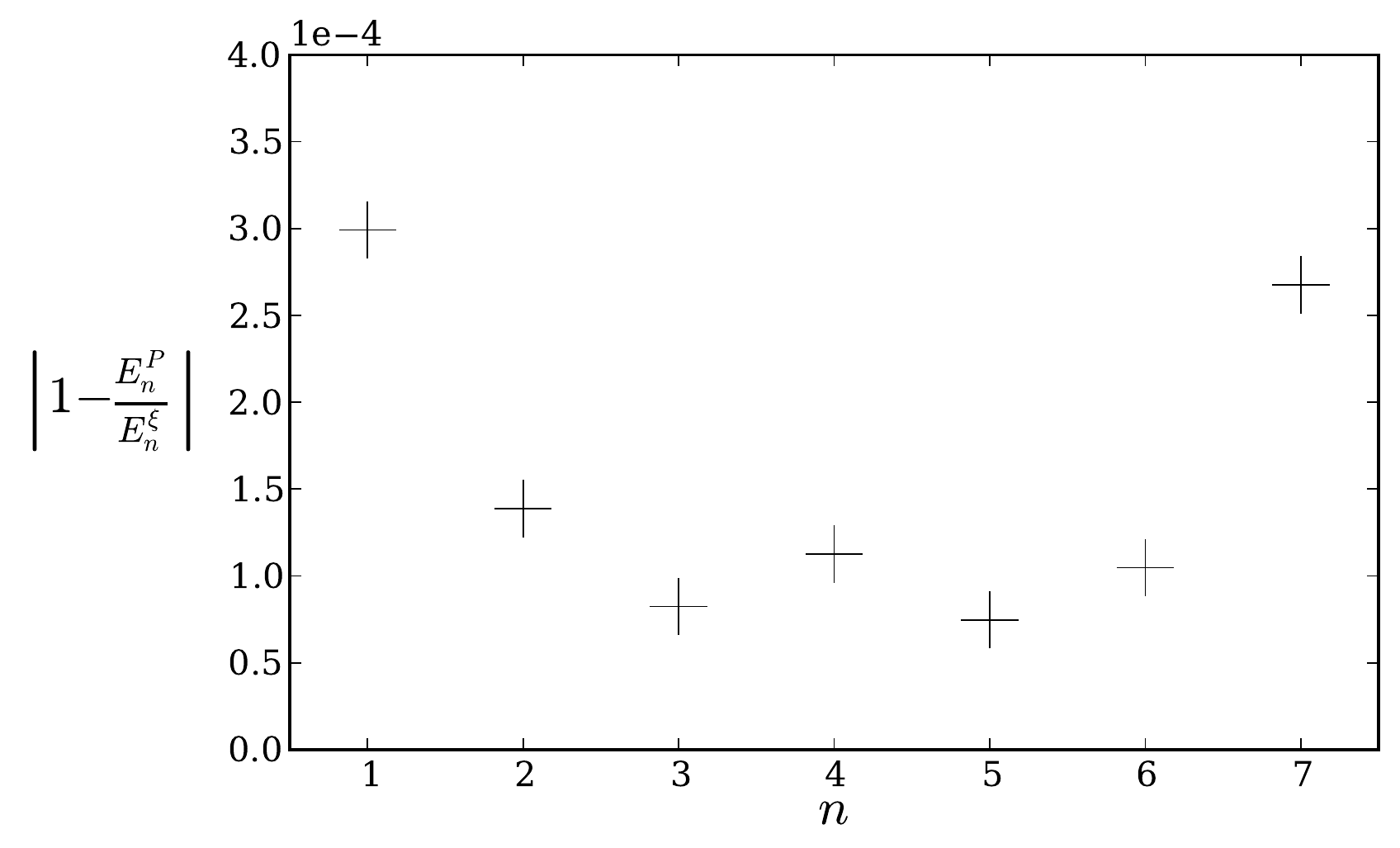}}
    \end{tabular}
    \caption{\small{A comparison between two methods of finding E-COSEBIs.
	      $E_n^{P}$ is calculated from \Eqt\eqref{eq:EnFourier},
	      while $E_n^{\xi}$ estimated from \Eqt\eqref{eq:EnReal}.
	      $E_n$ with $n=1-7$ are shown here for an angular range of $[1',400']$.}}
    \label{fig:EnCompare}
  \end{center}
\end{figure}

In \cite{Asgari12} we calculated COSEBIs numerically, assuming a perfect knowledge, i.e. a noise-free measurement, of the input quantities. 
We used \Eqt\eqref{eq:EnFourier} to find the E-mode COSEBIs which is more convenient to use for a theoretical analysis, 
since most theories provide us with an input power spectrum. 
However, in practice shear 2PCFs are more straightforward to measure, from which COSEBIs can then be inferred via
\Eqt\eqref{eq:EnReal}.
The first test therefore checks if the two equations \Eqt\eqref{eq:EnFourier} and \Eqt\eqref{eq:EnReal}
result in the same $E_n$ when calculated numerically assuming noise-free data. 
% In principle, the $E_n$ estimated 
% from \Eqt\eqref{eq:EnFourier} and \Eqt\eqref{eq:EnReal} should be identical. 
% However, since these integrals are estimated numerically and the fact that the integral in
% \Eqt\eqref{eq:EnFourier} has an infinite upper limit, the results may differ. 
For this test we choose an angular range of $[1',400']$ .
\fig\ref{fig:EnCompare} shows the residual ratio of $E_n$ numerically calculated from 
\Eqt\eqref{eq:EnFourier} and \Eqt\eqref{eq:EnReal} for $n=1-7$. 
As we can see in this figure, the values of $E_n$ from the two methods agree to better than $0.03\%$.
Figure\thinspace10 of \cite{Asgari12} shows the dependence of three cosmological parameters to the 
first 5 COSEBIs. From this figure we conclude that the small difference between $E_n$ from
\Eqt\eqref{eq:EnFourier} and \Eqt\eqref{eq:EnReal} is therefore indeed insignificant for our analysis
(for example, if $\sigma_8$ changes by $1\%$ then $E_1$ will change by about $2\%$).

\subsection{From Smooth Integration to Noisy Trapezoidal}
\label{app:trap}

In \App\ref{app:PvsKsi} we assumed a perfect knowledge of the 2PCFs 
over the angular range considered, a Gaussian integration method 
\citep[see][]{numerical} between two extrema of the integrand is employed 
to evaluate $E_n$ in both cases. In practice, however, we only 
have the values of 2PCFs in angular bins or at certain $\theta$ values. 
Consequently, we need to use a different integration routine to
evaluate $E_n$ from \Eqt\eqref{eq:EnReal} for real data. 
The most straightforward integration method is the trapezoidal method for a linearly binned data. 
In this section we determine how many linear angular bins are needed to reach 
a certain accuracy in determining $E_n$.

\begin{figure}
  \begin{center}
    \begin{tabular}{c}
      \resizebox{80mm}{!}{\includegraphics{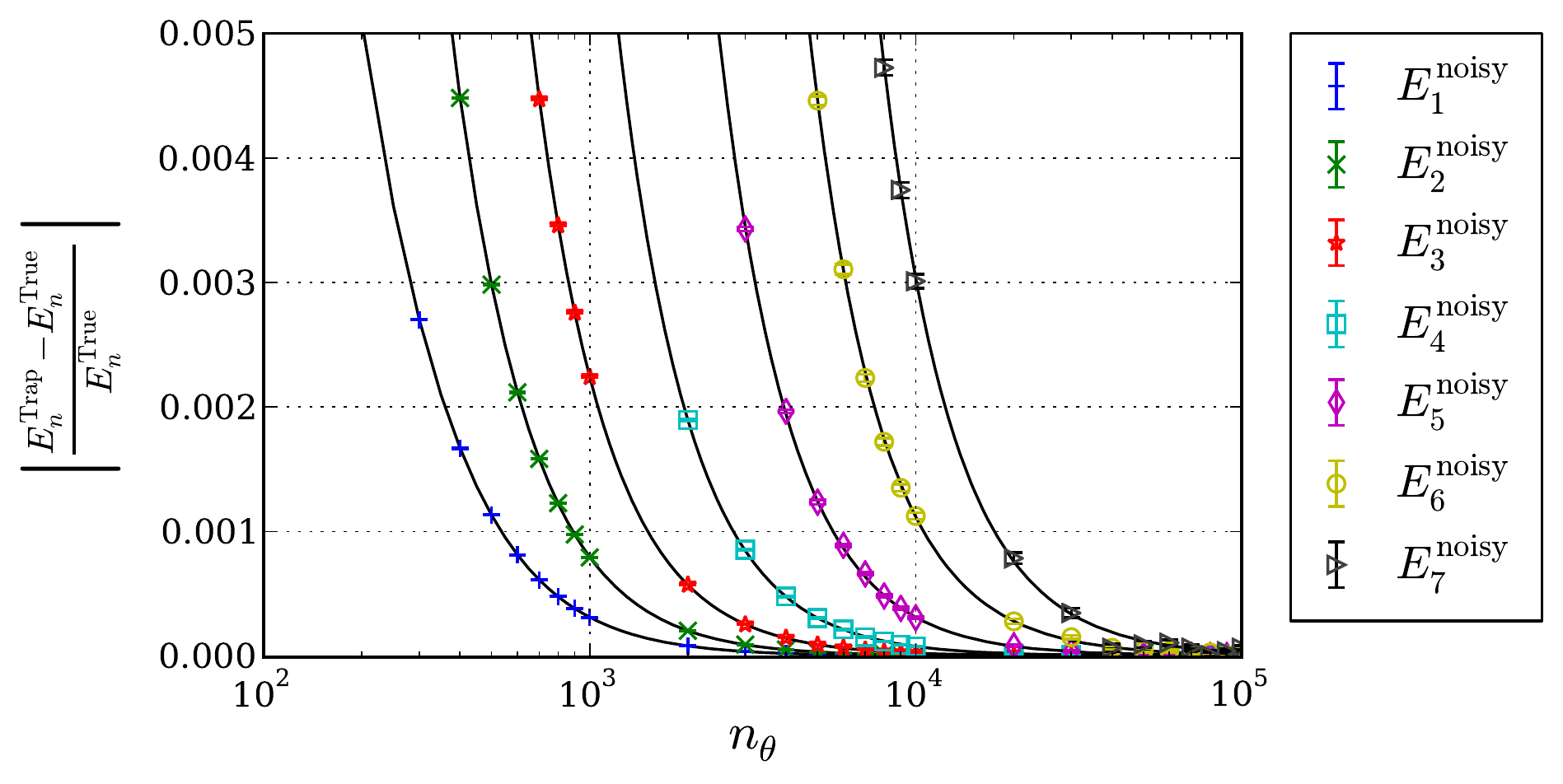}}
    \end{tabular}
    \caption{\small{The effect of noise on the estimated $E_n$ for the first 7 modes. 
		  Here we only consider uncorrelated noise between the angular bins. 
		  $n_\theta$ is the number of angular bins for the 2PCFs. 
		  The error bars are calculated from the variance between the noise realizations.
		  The black solid curves show the $E_n$ values without any added noise.} }
    \label{fig:EnTrap}
  \end{center}
\end{figure}

The solid curves in \fig\ref{fig:EnTrap} show the fractional deviation of  $E_n$ as a function of angular bins used in 
the trapezoidal integration assuming noise-free data. All the $E_n$ values are normalized by 
their true value, calculated from the convergence power spectrum using \Eqt\eqref{eq:EnFourier}. 
As can be seen in \fig\ref{fig:EnTrap} a larger number of angular bins 
are required for the higher COSEBIs modes, to reach the same accuracy. 
The reason for this behaviour is that, the $T_{\pm n}(\vartheta)$ functions have $n+1$ roots in their range 
of support and oscillate around them.
Consequently, the higher modes are more sensitive to the number 
of $\theta$ bins incorporated in their integral \citep[see][]{SEK10,Asgari12}.
Following \fig10 in \cite{Asgari12} we choose 
an accuracy of $0.5\%$ for $E_n$ which corresponds to an accuracy on $\sigma_8$ of $0.25\%$.
This enforces a lower limit of 10000 bins for, $E_7$, the highest COSEBIs mode we use in this analysis.

We next include the effects of noise on the estimated $E_n$. 
Since the number of bins needed to reach the accuracy desired is high, 
the shot noise term of the correlation function covariance dominates 
the other terms in this case. 
We will therefore choose to ignore other sources of noise for this test,
using only the shot noise in the covariance, at this stage. 
We can make a noisy $\xi_{\pm}(\theta)$ mock data set from
\begin{equation}
 \xi_{\pm}(\theta)^{\rm Noisy}=\xi_{\pm}(\theta)+N\times R\;,
\end{equation}
where 
\begin{equation}
 N=\sqrt{\frac{\sigma_\epsilon^4}
 {4\pi A \bar{n}^2\Delta\theta\:\theta}}\;,
\end{equation}
is the square root of the shot noise term in the 2PCFs covariance 
\citep[see][for example]{JoachimiSchneder08}, with
$\sigma_\epsilon=0.279$, the intrinsic ellipticity dispersion of the galaxies, 
$A=154$ deg$^2$, the area of the survey, $\bar{n}=11$ arcmin$^2$, the effective mean number density of galaxies and
$\Delta\theta$ the width of the angular bins.
$R$ is a randomly generated number from a Gaussian distribution with a variance of 1 and a mean of 0. 
With the above definitions the covariance of $\xi_{\pm}(\theta)^{\rm Noisy}$ 
is equal to the desired covariance. 
The symbols in \fig\ref{fig:EnTrap} show the ensemble average estimate of $E_n$ from 
50 $\xi_{\pm}(\theta)^{\rm Noisy}$ realizations with respect to the number of angular bins. 
The errors shown are the standard deviation of the mean value of $E_n$ over all the realizations. 
The presence of the random errors do not change the conclusions drawn from the previous test. 
By comparing the curves and the symbols in \fig\ref{fig:EnTrap} we can 
also conclude that the random noise due to the intrinsic ellipticity dispersion of galaxies 
does not bias the estimation of COSEBIs.

\begin{figure}
  \begin{center}
    \begin{tabular}{c}
      \resizebox{80mm}{!}{\includegraphics{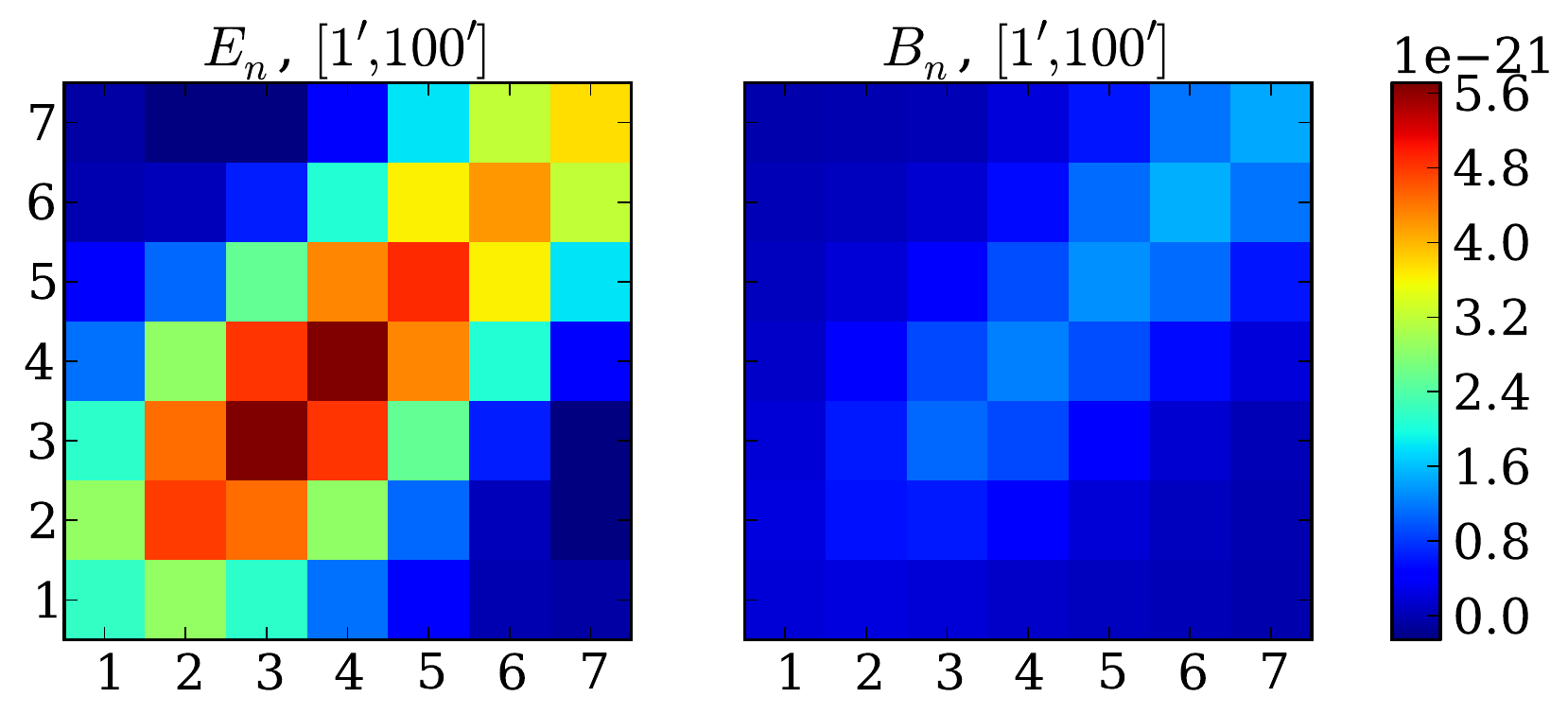}} \\
       \resizebox{80mm}{!}{\includegraphics{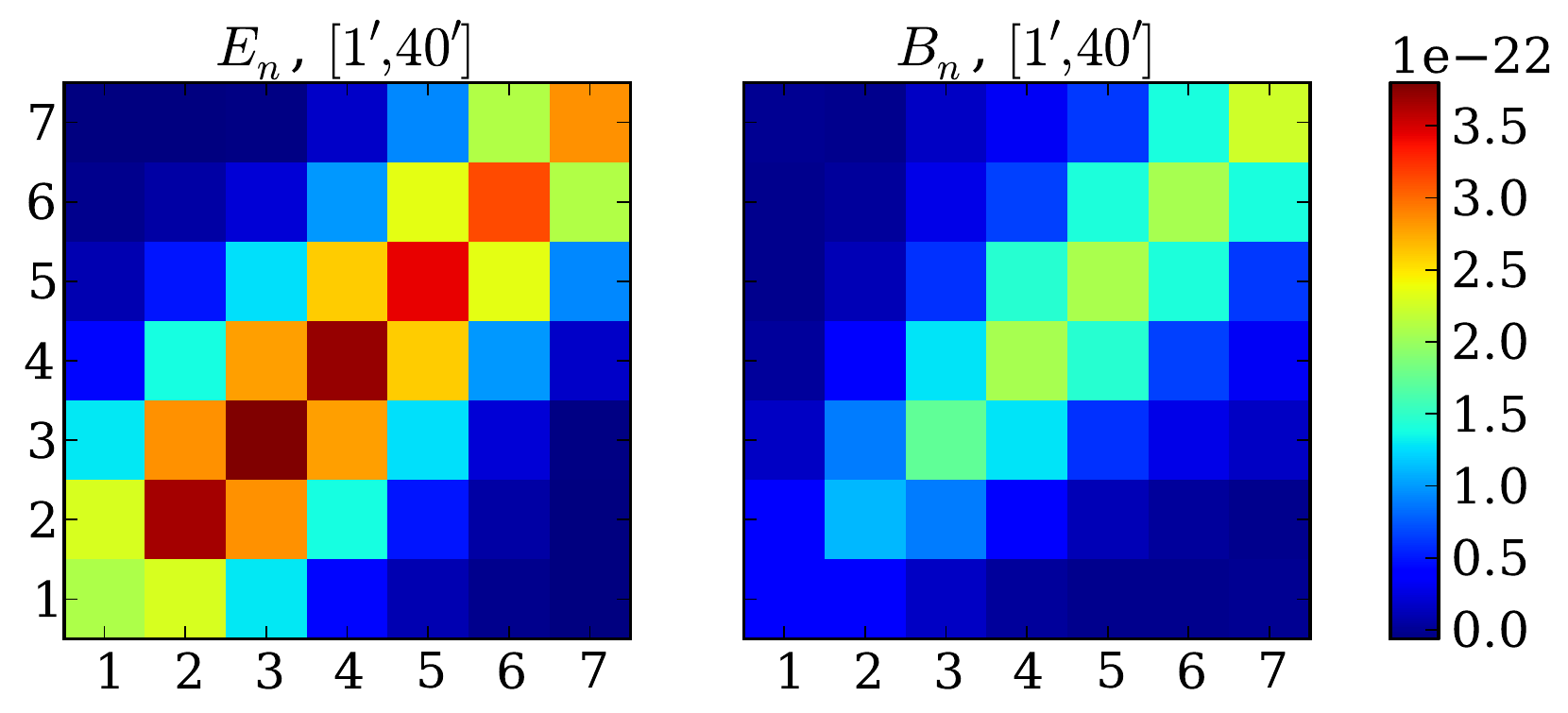}} \\
      \resizebox{80mm}{!}{\includegraphics{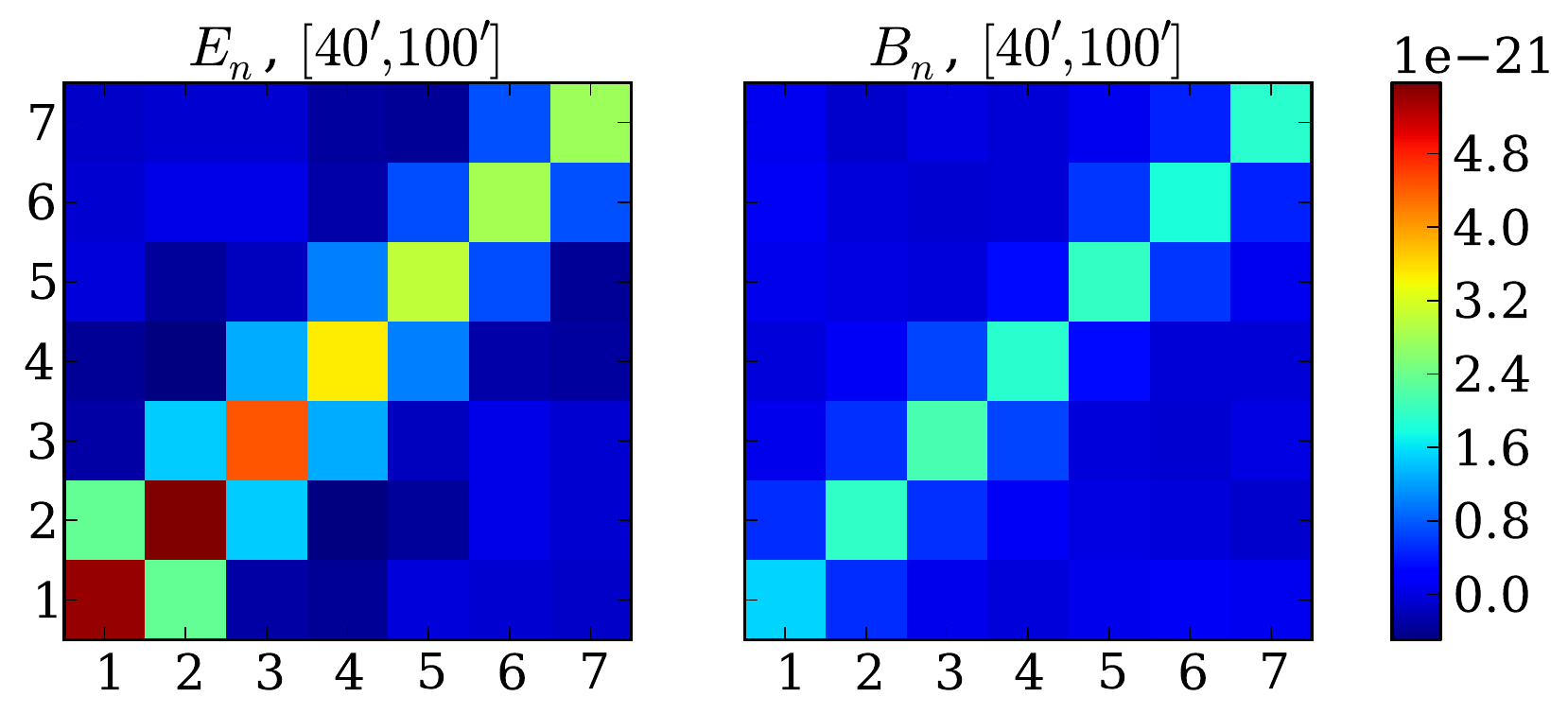}} 
    \end{tabular}
    \caption{\small{Covariance matrices of COSEBIs for a single redshift distribution. 
    Three angular ranges, $[1',100']$, $[1',40']$ and $[40',100']$ are considered here. 
    The x/y-axis show the COSEBIs mode considered.
    The COSEBIs are less correlated for $[40',100']$ compared to the other cases.}}
    \label{fig:CovSim1bin}
  \end{center}
\end{figure}

\subsection{Simulations: Covariance Matrix Estimation}
\label{app:SLICS}

% $A_{\rm CFHTLenS}=94.563993$ deg$^2$.
In this section we determine COSEBIs on the mock data from the SLICS simulations which resemble the CFHTLenS data. 
Here we also show the covariance matrices which are used in the main analysis 
and are estimated from the simulated data.
In this paper we use the second version of the CFHTLenS CLONE catalogue\footnote{The first version is available on {\tt www.cfhtlens.org}}, 
which is based on the SLICS N-body simulations \citep[see][for details]{Harnois15} 
and consist of 497 independent lines of sight, 60 square degrees each. 
These mock catalogues are specifically made for the CFHTLenS data, taking into account its
redshift distribution. Furthermore, source clustering effects are included in these catalogues.
The limited box size of the simulations dictates the maximum scale that can be trusted. 
In addition, the resolution of the simulations put limits on the small scales. 
Combining this with the fact that on small scales baryonic effects (not included in the simulations) 
 become important, we limit our minimum angular range as well
\citep[see][for the effects different baryon feedback models have on structure formation and 
$\xi_{\pm}$]{Semboloni11}. Hence, we choose to only use scales in $[1', 100']$ in our analysis. 

% We will see in the following sections that the signal is 
% very small for $[20',40']$.
% % , hence conclude that higher angular scales have a negligible contribution
% % to the analysis. 
% \cite{Heymans13} summarize the $\sigma_8(\Omega_{\rm m}/0.27)^{\alpha}$ constraints
% for several cases (see Table 2 in their paper).  
% This parametrisation characterizes the degeneracy between $\sigma_8$ and $\Omega_m$.
% They find that using larger angular scales increases the constraints on this parameter. 
% However, in all of the cases considered by \cite{Heymans13}, other cosmological parameters
% are not fixed to their fiducial values and are marginalized over. 
% Consequently, higher angular ranges add more information for those cases, while
% they do not have a significant impact in the analysis presented here.

Since the simulated covariance is different from the Gaussian random noise we used in \App\ref{app:trap}
we repeat the angular bin versus measured $E_n$ test using the simulated data.
The covariance of the COSEBIs is defined as,
\begin{equation}
 \tens{C}_{mn}\equiv \langle E_m E_n\rangle-\langle E_m \rangle \langle E_n\rangle\;,
\end{equation}
where $\langle E_n \rangle$ is the expectation value of $E_n$. 
The covariance is estimated from the simulations via,
\begin{equation}
 \tens{C}_{mn}=\frac{1}{N-1}\sum_{i=1}^N (E^i_m-\langle{E}_m\rangle)(E^i_n-\langle{E}_n\rangle)\;,
\end{equation}
where 
\begin{equation}
 \langle{E}_n\rangle=\frac{1}{N}\sum_{i=1}^N E^i_n\;,
\end{equation}
is the mean $E_n$ over all lines-of-sight of the $N=497$ simulated fields.
% 
% The binning is linear which is more convenient for the integral in \Eqt\eqref{eq:EnReal}.

We find similar conclusions from repeating the bin size exercise.  
Nevertheless, for the rest of the analysis we choose to use $4\times10^5$ linear bins in $[1', 100']$,
which is larger than the threshold we found in the previous section.
With a narrower angular binning scheme the number of galaxies in each bin decreases. 
Hence the $\xi_\pm$ estimate is noisier. In this analysis we have made sure that all the bins are
populated with galaxies. Increasing the number of 2PCFs angular bins cannot reduce the accuracy of the estimated COSEBIs, 
as long as all the angular bins are populated with pairs of galaxies.
% a similar band matrix form, however non-Gaussian effects are important here, since the matter power
% spectrum for these scales becomes non-linear which causes mode mixing. 
% This effect differentiates between the theory and simulations.

% \begin{figure}
%   \begin{center}
%     \begin{tabular}{c}
%       \resizebox{70mm}{!}{\includegraphics{plots/CloneCOSEBIsvsNTheta.pdf}}
%     \end{tabular}
%     \caption{\small{$E_n$ versus the number of angular bins, $N_{\theta}$, for SLICS simulations.
%     The angular range here is $[1',40']$. 
%     The error bars come from the covariance measured from the simulations.
%     We can see here that using more than about $10^5$ angular bins does not contribute to 
%     a better convergence to the true value of $E_n$, since the fluctuations in the measurements
%     are within the statistical errors.}}
%     \label{fig:nTrapSims}
%   \end{center}
% \end{figure}

\begin{figure*}
  \begin{center}
    \begin{tabular}{c}
      \resizebox{180mm}{!}{\includegraphics{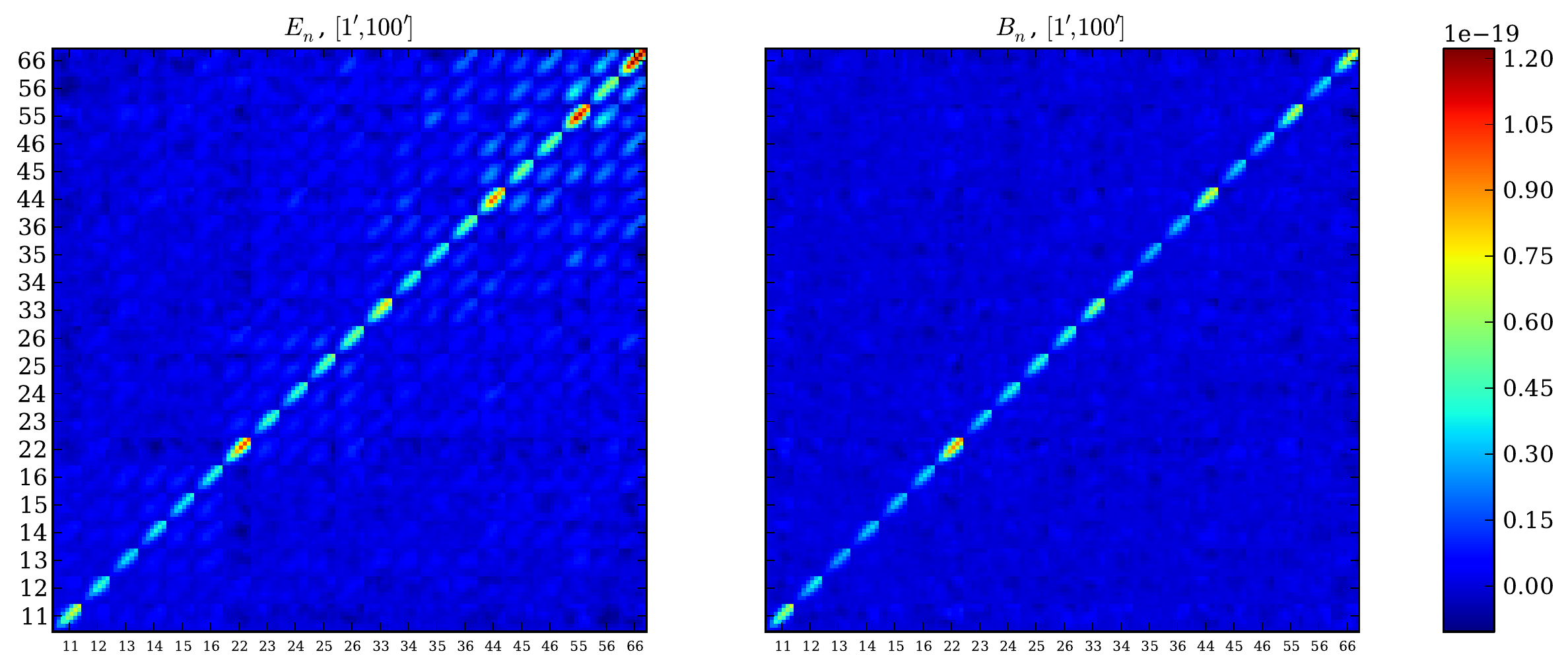}} \\
    \end{tabular}
    \caption{\small{Covariance matrices of COSEBIs for 6 redshift bins. 
    Three angular ranges, $[1',100']$, $[1',40']$ and $[40',100']$ are considered here.
    The x/y-axis show the redshift bin combination, for example '13' 
    means redshift bins 1 and 3 are relevant. There are 7 COSEBIs modes for each combination.}}
    \label{fig:CovSim6bin}
  \end{center}
\end{figure*}

\begin{figure*}
  \begin{center}
    \begin{tabular}{c}
      \resizebox{180mm}{!}{\includegraphics{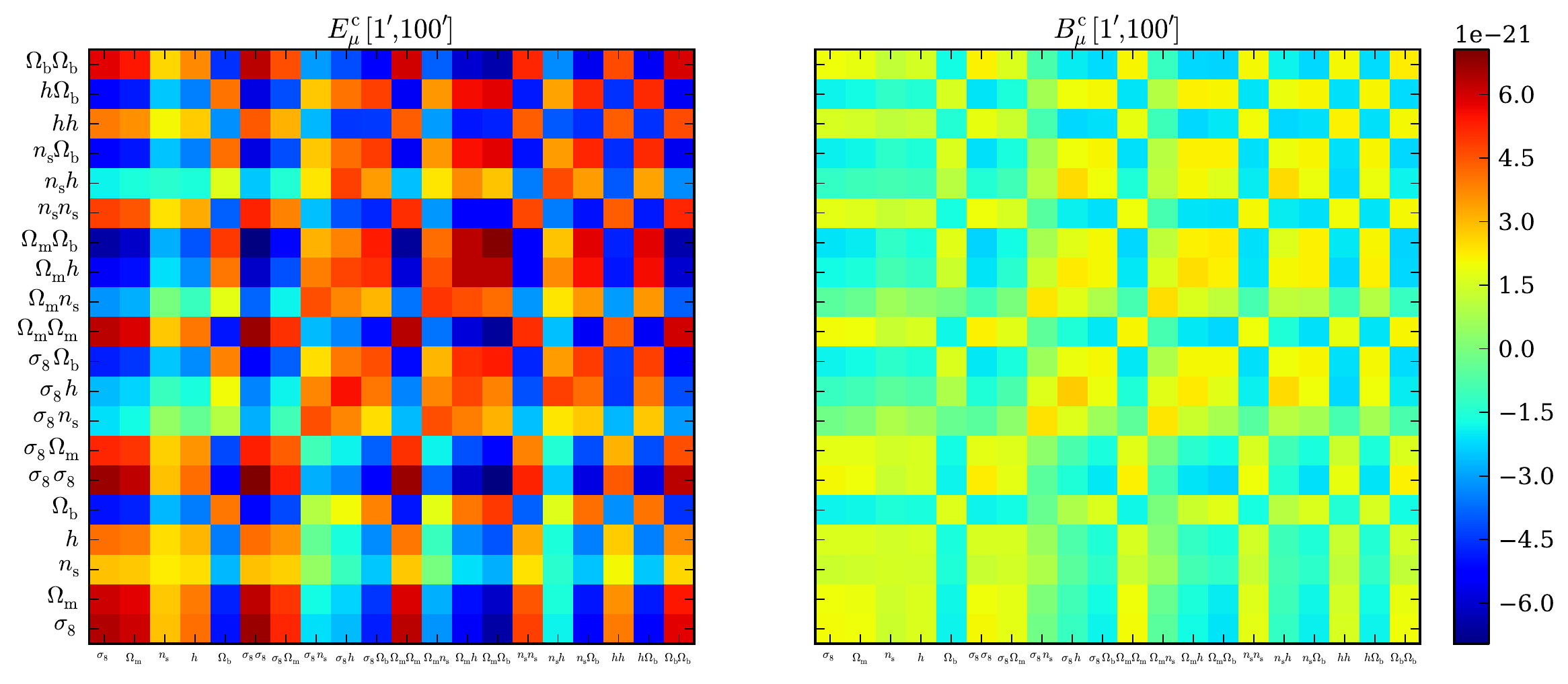}}  
    \end{tabular}
    \caption{\small{Covariance matrices of CCOSEBIs for 6 redshift bins estimated from SLICS simulations. 
    The angular range $[1',100']$ is considered here. 
    The left and right panels show the covariances for the E-modes and B-modes respectively.
    The CCOSEBIs are linear combinations of COSEBIs. The CCOSEBIs modes are denoted by the parameter(s) 
    that they are most sensitive to. A comparison of the
    number of elements in the COSEBIs covariance for 6 redshift bins in \fig\ref{fig:CovSim6bin} 
    and this figure shows the significance of this compression method.}}
    \label{fig:CovSimCCOSEBIs}
  \end{center}
\end{figure*}

\begin{figure*}
  \begin{center}
    \begin{tabular}{c}
      \resizebox{85mm}{!}{\includegraphics{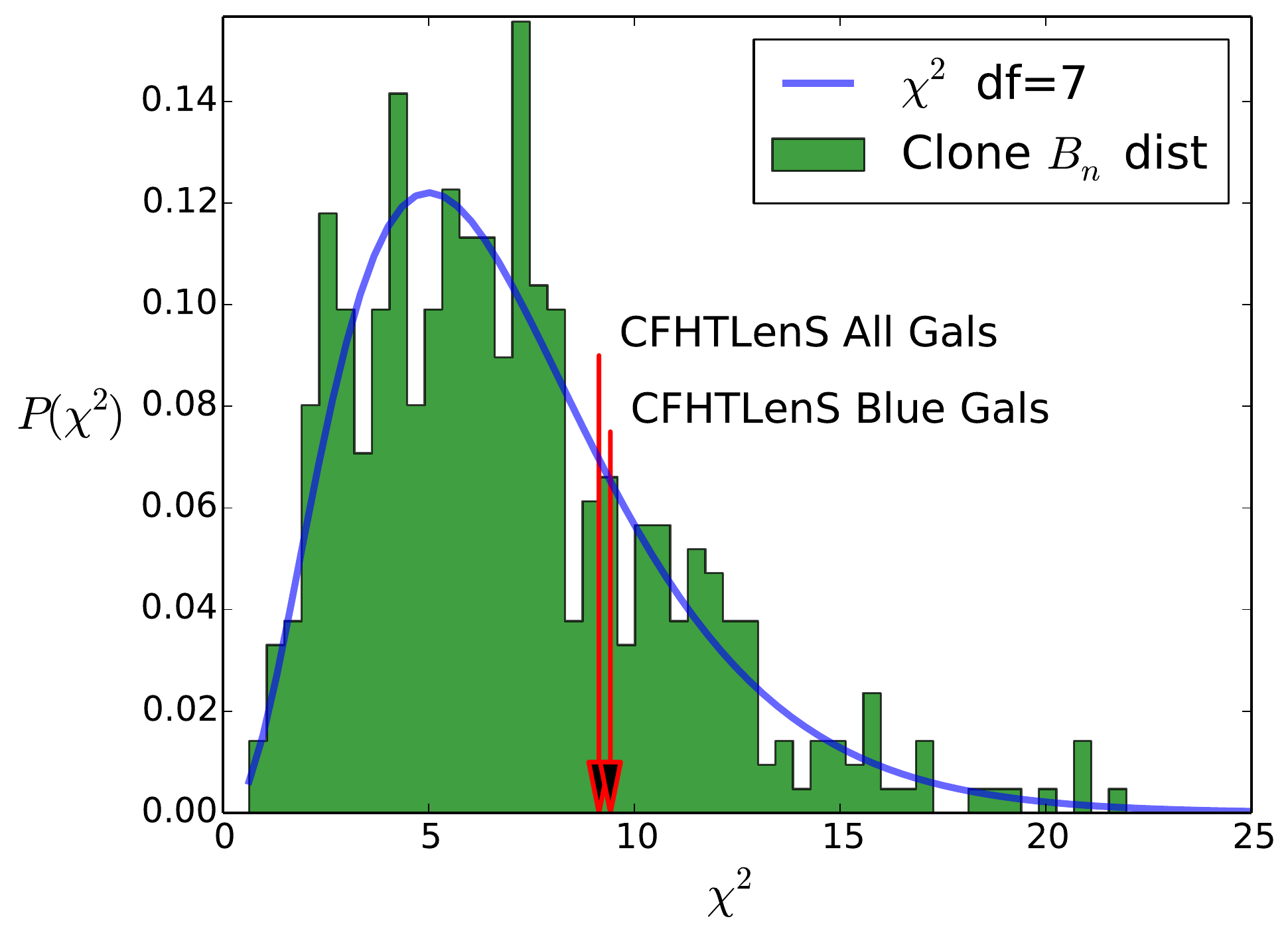}}
      \resizebox{85mm}{!}{\includegraphics{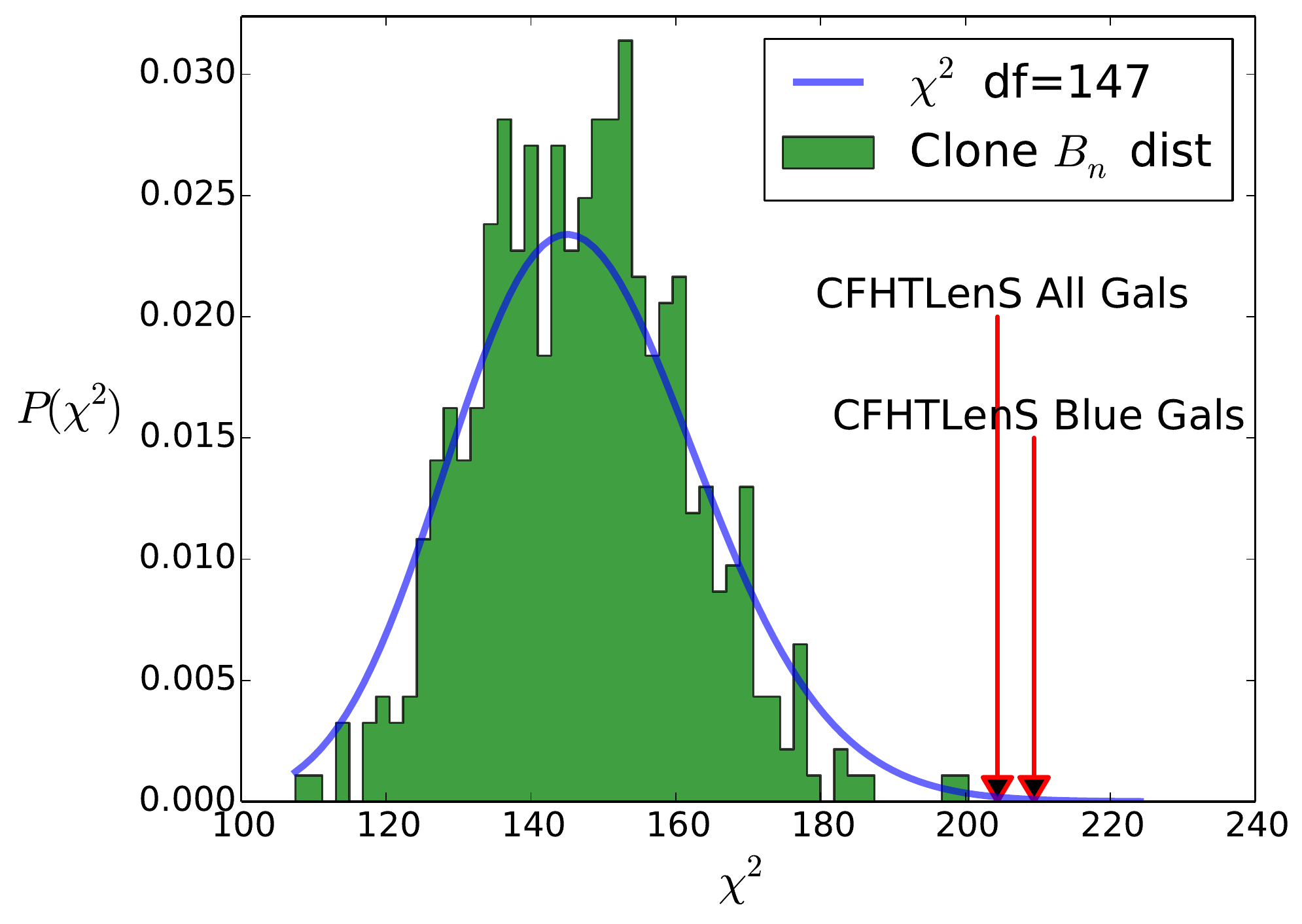}}
    \end{tabular}
    \caption{\small{$\chi^2$ distribution of $B_n$ assuming a zero B-modes model for the Clone simulations 
    (green histogram). The left plot corresponds to a single redshift distribution with 7 $B_n$ modes,
    while the right corresponds to 6 redshift bins with 7 $B_n$ modes resulting in 147 modes in total.
    The blue curves show the theoretical $\chi^2$ distribution with the given degrees-of-freedom, df. 
    The arrows show the $\chi^2$ values for the B-modes in $[1',100']$ 
    range in the CFHTLenS data for the corresponding cases. }}
    \label{fig:ChiSDist}
  \end{center}
\end{figure*}

The COSEBIs covariance for a single redshift distribution from the simulations is shown in 
\fig\ref{fig:CovSim1bin}, for the three angular ranges $[1',100']$, $[1',40']$ and $[40',100']$. 
The right panels show the covariance for $E_n$ while the left panels show the same for $B_n$. 
All the covariances are scaled by a factor of $60\times 497/94.564$ to correspond to the effective CFHTLenS area,
where we have 497 mock fields, 60 deg$^2$ each. The effective area of CFHTLenS passed fields is 94.564 deg$^2$.

In the simulated catalogues a mock best fit value of each galaxy's redshift, $z_{\rm B}$, is given, which is in general different 
from its spectroscopic value. To mimic the real data we use $z_{\rm B}$ to choose which redshift bin 
a galaxy belongs to. 
% while using the underlying redshift distribution of all the galaxies in each
% tomographic bin to find the theory values for our observables. 
Whilst tomographic bins have no overlap in $z_{\rm B}$, this is not the case for the underlying 
true redshift distribution.

\fig\ref{fig:CovSim6bin} shows the covariance matrices for $[1',100']$ with the six redshift bins 
in \tab\ref{tab:CFHTLenSNeff}. Each block in the covariance matrix has $7^2$ elements corresponding to 
a combination of redshift bin pairs. The x/y-axis in the plots show the redshift bin pairs considered. 
In total the covariance has $147^2$ elements. 
The left and right panels show the covariance matrix for the $E_n$ and $B_n$, respectively. 
In \fig\ref{fig:CovSim6bin} we see that for all cases, the value of the covariance drops for the off-diagonal elements.
Although not shown here for the tomographic case, for both redshift binning 
cases the $[40',100']$ range has a more diagonal covariance.
This is due to the fact that non-Gaussian effects are less important for this angular range. 
Therefore, any analysis which only uses this angular range is less likely to be biased because of poor 
modelling of non-linear scales. However, the cosmic shear information in this angular range 
is significantly lower than that of the lower angular scales. 

\fig\ref{fig:CovSimCCOSEBIs} shows the covariance matrices for the
E-mode and B-mode CCOSEBIs for six redshift bins and $\theta\in[1',100']$,
measured from the SLICS simulations. The x-/y-axis show the CCOSEBIs modes for which the covariance is shown. 
In this work we chose five cosmological parameters, 
$\sigma_8$, $\Omega_{\rm m}$, $n_s$, $h$ and $\Omega_{\rm b}$. This means that
we have 5 first order CCOSEBIs which depend on the covariance and the first order derivatives of the COSEBIs
with respect to the parameters, 
and 15 second order CCOSEBIs which depend on the covariance and the second order derivatives of the COSEBIs
with respect to the parameters. 
Hence we show the CCOSEBIs modes by the parameters with respect to which the compression is made. 
In total for 5 parameters there are 20 CCOSEBIs modes irrespective of the number of redshift bins. Hence,
for the case of 6 redshift bins we have compressed 147 parameters to only 20 and reduced the size of the covariance
substantially, as can be seen by comparing \fig\ref{fig:CovSim6bin} and \fig\ref{fig:CovSimCCOSEBIs}.
The cross-covariance between the CCOSEBIs modes is relatively high for some of the cases. 
This is due to the fact that the CCOSEBIs modes are based on cosmological parameters which can have large degeneracies.
For example, the $\Omega_{\rm m}$ and $\sigma_8$ which have a large degeneracy in cosmic shear analysis, also show a large cross-covariance.

\subsection{B and E-mode analysis of mocks}

The simulated mock catalogues should be B-mode free, providing an opportunity 
to explore how random ellipticity noise can affect the measured
B-modes.
% We also checked our simulations for B-modes using COSEBIs. The simulations in principle should be B-mode free. 
% However, the random ellipticity noise added to the galaxy catalogues can produce some artificial B-modes. 
For each line-of-sight we measure all $B_n$ values and determine the $\chi^2$ values
for $B_n=0$,
\begin{equation}
 \chi^2_{B}=\boldsymbol{B}^{\rm t}\tens{C}^{-1}\boldsymbol{B}\;,
\end{equation}
where $\boldsymbol{B}$ is the vector and $\tens{C}$ is the covariance matrix of $B_n$. 
\fig\ref{fig:ChiSDist} shows the distribution of the $\chi^2_{B}$ for the 497 SLICS simulations (green histograms).
The left panel belong to the single redshift case, whereas the tomographic case is shown in the right panel.
Since, in this study we used 7 COSEBIs modes the degrees-of-freedom for the single redshift bin case is 7,
while for the tomographic case it is 147. The blue solid curves show the theoretical $\chi^2$ distribution
for a given degrees-of-freedom, and they match the histograms\footnote{We checked this using a Kolmogorov–Smirnov test.}.
Consequently, we conclude that the B-modes in the simulations are statistically consistent with zero
and provide a $\chi^2$ distribution with which to compare the real data.
The red arrows show the value of the B-modes $\chi^2$ corresponding to the CFHTLenS data
with $\theta\in[1',100']$. 
We can see that the blue galaxies show a more significant B-mode signal compared to all galaxies.  
Furthermore, the $\chi^2$ values of the CFHTLenS data for the tomographic case are well beyond what is expected 
from the simulations. We also calculated the CCOSEBIs from the simulations and confirmed that they follow a $\chi^2$
distribution with 20 degrees-of-freedom.

The E-mode COSEBIs were also optimized and tested using the same set of catalogues. 
We performed a blind analysis of the mocks to test if the input cosmology of the simulations can be recovered. 
The $\chi^2$ distributions for the $E_n-E_n^{th}$, where $E_n$ is estimated from the simulations and  $E_n^{th}$
is its expected theory value, is very similar to the $\chi^2$ distributions of the $B_n$ (\fig\ref{fig:ChiSDist}), 
hence we do not show them here. The CFHTLenS data analysis was then carried out without any changes to the pipelines.

%%%%%%%%%%%%%%%%%%%%%%%%%%%%%%%%%%%%%%%%%%%%%%%%%%
\section{$\chi^2$ values and degrees-of-freedom}

Comparing raw $\chi^2$ values can be misleading for two reasons: firstly a $\chi^2$ distribution is asummetric
and secondly this distribution highly depends on the degrees-of-freedom associated with the $\chi^2$ value.
However, in \tab\ref{tab:DOF} we provide the degrees-of-freedom for each element in \tab\ref{tab:PvalueAll} and the 
$\chi^2$ values to give the readers the opportunity to perform their own interpretation of the data.

\begin{table*}
\caption{\small{The number of degrees-of-freedom of the $\chi^2$ value for each entry in \tab\ref{tab:PvalueAll}.}}
\begin{center}
\begin{adjustbox}{max width=\textwidth}

\begin{tabular}{ c  c  c  c  c  c  c  c  c  c  c  c  c  c  c  c }
&&&&\multicolumn{ 6}{ c}{ COSEBIs}&\multicolumn{ 6}{ c}{ CCOSEBIs}\\ \hline
$\theta$ range & Galaxies & z-bins & $n_{\rm max}$
&\multicolumn{ 2}{ c}{$B_n$} &\multicolumn{ 2}{ c}{$E_n^{\rm CFHTLenS}$} & \multicolumn{ 2}{ c}{$E_n^{\rm Planck}$} 
 & \multicolumn{ 2 }{ c }{$B^{\rm c}$ } & \multicolumn{ 2}{ c}{$E^{\rm c, CFHTLenS}$}& \multicolumn{ 2}{ c}{$E^{\rm c, Planck}$} 
 \\ \hline
&&&& DOF & $\chi^2$ & DOF & $\chi^2$ & DOF & $\chi^2$ & DOF & $\chi^2$ & DOF & $\chi^2$ & DOF & $\chi^2$ \\ \hline
\multirow{8}{*}{$[1'-100']$}& \multirow{4}{*}{All}& \multirow{2}{*}{1}&2 &$ 2$ &$ 1.60$ &$ 2$ &$ 3.67$ &$ 2$ &$ 2.15$ &$-$ &$-$ &$-$ &$-$ &$-$ &$-$
\\ \cline{4-16}
& &&7 &$ 7$ &$ 9.29$ &$ 7$ &$ 14.71$ &$ 7$ &$ 9.73$ &$-$ &$-$ &$-$ &$-$ &$-$ &$-$
\\ \cline{4-16}

 \cline{3-16}
& & \multirow{2}{*}{6}&2 &$ 42$ &$ 82.85$ &$ 42$ &$ 80.16$ &$ 42$ &$ 90.06$ &$ 20$ &$ 34.70$ &$ 20$ &$ 31.05$ &$ 20$ &$ 39.37$
\\ \cline{4-16}
& &&7 &$ 147$ &$ 291.33$ &$ 147$ &$ 322.10$ &$ 147$ &$ 335.59$ &$ 20$ &$ 15.32$ &$ 20$ &$ 51.11$ &$ 20$ &$ 53.41$
\\ \cline{4-16}

 \cline{3-16}

 \cline{2-16}
& \multirow{4}{*}{Blue}& \multirow{2}{*}{1}&2 &$ 2$ &$ 3.11$ &$ 2$ &$ 5.43$ &$ 2$ &$ 2.10$ &$-$ &$-$ &$-$ &$-$ &$-$ &$-$
\\ \cline{4-16}
& &&7 &$ 7$ &$ 9.57$ &$ 7$ &$ 23.20$ &$ 7$ &$ 18.24$ &$-$ &$-$ &$-$ &$-$ &$-$ &$-$
\\ \cline{4-16}

 \cline{3-16}
& & \multirow{2}{*}{6}&2 &$ 42$ &$ 90.28$ &$ 42$ &$ 66.91$ &$ 42$ &$ 77.73$ &$ 20$ &$ 32.31$ &$ 20$ &$ 24.55$ &$ 20$ &$ 31.77$
\\ \cline{4-16}
& &&7 &$ 147$ &$ 298.54$ &$ 147$ &$ 285.77$ &$ 147$ &$ 295.41$ &$ 20$ &$ 20.67$ &$ 20$ &$ 29.42$ &$ 20$ &$ 34.65$
\\ \cline{4-16}

 \cline{3-16}

 \cline{2-16}

\cline{1-16}
\multirow{8}{*}{$[1'-40']$}& \multirow{4}{*}{All}& \multirow{2}{*}{1}&2 &$ 2$ &$ 0.60$ &$ 2$ &$ 6.97$ &$ 2$ &$ 1.36$ &$-$ &$-$ &$-$ &$-$ &$-$ &$-$
\\ \cline{4-16}
& &&7 &$ 7$ &$ 4.43$ &$ 7$ &$ 11.79$ &$ 7$ &$ 6.41$ &$-$ &$-$ &$-$ &$-$ &$-$ &$-$
\\ \cline{4-16}

 \cline{3-16}
& & \multirow{2}{*}{6}&2 &$ 42$ &$ 66.64$ &$ 42$ &$ 80.64$ &$ 42$ &$ 85.12$ &$ 20$ &$ 24.98$ &$ 20$ &$ 35.09$ &$ 20$ &$ 39.45$
\\ \cline{4-16}
& &&7 &$ 147$ &$ 275.17$ &$ 147$ &$ 290.76$ &$ 147$ &$ 307.95$ &$ 20$ &$ 18.21$ &$ 20$ &$ 43.23$ &$ 20$ &$ 47.66$
\\ \cline{4-16}

 \cline{3-16}

 \cline{2-16}
& \multirow{4}{*}{Blue}& \multirow{2}{*}{1}&2 &$ 2$ &$ 0.86$ &$ 2$ &$ 8.17$ &$ 2$ &$ 0.26$ &$-$ &$-$ &$-$ &$-$ &$-$ &$-$
\\ \cline{4-16}
& &&7 &$ 7$ &$ 8.88$ &$ 7$ &$ 22.03$ &$ 7$ &$ 16.46$ &$-$ &$-$ &$-$ &$-$ &$-$ &$-$
\\ \cline{4-16}

 \cline{3-16}
& & \multirow{2}{*}{6}&2 &$ 42$ &$ 66.43$ &$ 42$ &$ 69.16$ &$ 42$ &$ 72.39$ &$ 20$ &$ 16.93$ &$ 20$ &$ 23.36$ &$ 20$ &$ 26.25$
\\ \cline{4-16}
& &&7 &$ 147$ &$ 226.22$ &$ 147$ &$ 287.05$ &$ 147$ &$ 304.01$ &$ 20$ &$ 31.20$ &$ 20$ &$ 28.35$ &$ 20$ &$ 32.54$
\\ \cline{4-16}

 \cline{3-16}

 \cline{2-16}

\cline{1-16}
\multirow{8}{*}{$[40'-100']$}& \multirow{4}{*}{All}& \multirow{2}{*}{1}&2 &$ 2$ &$ 11.01$ &$ 2$ &$ 6.47$ &$ 2$ &$ 5.40$ &$-$ &$-$ &$-$ &$-$ &$-$ &$-$
\\ \cline{4-16}
& &&7 &$ 7$ &$ 22.79$ &$ 7$ &$ 13.74$ &$ 7$ &$ 12.68$ &$-$ &$-$ &$-$ &$-$ &$-$ &$-$
\\ \cline{4-16}

 \cline{3-16}
& & \multirow{2}{*}{6}&2 &$ 42$ &$ 85.72$ &$ 42$ &$ 73.14$ &$ 42$ &$ 71.48$ &$ 20$ &$ 33.30$ &$ 20$ &$ 42.92$ &$ 20$ &$ 41.92$
\\ \cline{4-16}
& &&7 &$ 147$ &$ 387.35$ &$ 147$ &$ 376.98$ &$ 147$ &$ 375.72$ &$ 20$ &$ 34.68$ &$ 20$ &$ 41.61$ &$ 20$ &$ 40.31$
\\ \cline{4-16}

 \cline{3-16}

 \cline{2-16}
& \multirow{4}{*}{Blue}& \multirow{2}{*}{1}&2 &$ 2$ &$ 10.54$ &$ 2$ &$ 3.79$ &$ 2$ &$ 3.10$ &$-$ &$-$ &$-$ &$-$ &$-$ &$-$
\\ \cline{4-16}
& &&7 &$ 7$ &$ 22.29$ &$ 7$ &$ 14.72$ &$ 7$ &$ 14.11$ &$-$ &$-$ &$-$ &$-$ &$-$ &$-$
\\ \cline{4-16}

 \cline{3-16}
& & \multirow{2}{*}{6}&2 &$ 42$ &$ 79.81$ &$ 42$ &$ 62.68$ &$ 42$ &$ 62.03$ &$ 20$ &$ 49.61$ &$ 20$ &$ 29.42$ &$ 20$ &$ 28.83$
\\ \cline{4-16}
& &&7 &$ 147$ &$ 391.02$ &$ 147$ &$ 341.97$ &$ 147$ &$ 341.87$ &$ 20$ &$ 48.61$ &$ 20$ &$ 27.27$ &$ 20$ &$ 26.84$
\\ \cline{4-16}

 \cline{3-16}

 \cline{2-16}

\cline{1-16}
\end{tabular}

\end{adjustbox}
\end{center}
\label{tab:DOF}
\end{table*}

% Don't change these lines
\bsp	% typesetting comment
\label{lastpage}

\end{document}